\documentclass{article}

\usepackage{amssymb}
\usepackage{amsmath,amsfonts,amstext,amscd,amsbsy,bm}
\usepackage{algorithm}
\usepackage{algorithmic}
\usepackage{graphicx}
\usepackage{epsfig}
\usepackage{subfigure}
\usepackage{setspace}
\usepackage{lineno}
\usepackage{natbib}
\usepackage{multirow}
\usepackage{epstopdf}
\usepackage{relsize}

\usepackage{url}

\def\bC{\boldmath{C}}

\def\bbeta{\boldmath{\beta}}

\def\bu{\boldmath{u}}
\def\bc{\boldmath{c}}

\def\bX{\boldmath{X}}
\def\bZ{\boldmath{Z}}
\def\bzero{\boldsymbol{0}}

\def\smhalf{{\textstyle{\frac{1}{2}}}}
\def\aeps{a_{\varepsilon}}

\def\sigeps{\sigma_{\varepsilon}}

\def\aeps{a_{\varepsilon}}

\def\bc{\boldsymbol{c}}

\def\bh{\boldsymbol{h}}

\def\bu{\boldsymbol{u}}

\def\by{\boldsymbol{y}}

\def\bC{\boldsymbol{C}}

\def\bG{\boldsymbol{G}}

\def\bI{\boldsymbol{I}}

\def\bM{\boldsymbol{M}}

\def\bX{\boldsymbol{X}}

\def\bZ{\boldsymbol{Z}}

\def\bbeta{\boldsymbol{\beta}}

\def\bzero{\boldsymbol{0}}

\def\Aeps{A_{\varepsilon}}

\def\smhalf{{\textstyle{\frac{1}{2}}}}

\def\bmu{\boldsymbol{\mu}}
\def\bSigma{\boldsymbol {\Sigma}}
\def\punder{\underline{p}}
\def\simind{\stackrel{{\tiny \mbox{ind.}}}{\sim}}

\def\btheta{\boldsymbol{\theta}}

\def\auell{a_{u\ell}}
\newcommand{\ignore}[1]{}

\begin{document}

\title{\bf Real-time semiparametric regression for distributed data sets}
\author{\\
Jan Luts\thanks{Jan Luts (E-mail: \textit{jan.luts@uts.edu.au}; Tel.: +61 2 9514 2267; Fax: +61 2 9514 2260)} ,\\
School of Mathematical Sciences, University of Technology Sydney\\
\\}
\maketitle

\begin{center}
\textbf{Abstract}
\end{center}

This paper proposes a method for semiparametric regression analysis of large-scale data which are distributed over multiple hosts. This enables modeling of nonlinear relationships and both the batch approach, where analysis starts after all data have been collected, and the real-time setting are addressed. The methodology is extended to operate in evolving environments, where it can no longer be assumed that model parameters remain constant over time. Two areas of application for the methodology are presented: regression modeling when there are multiple data owners and regression modeling within the MapReduce framework. A website, \texttt{realtime-semiparametric-regression.net}, illustrates the use of the proposed method on United States domestic airline data in real-time.

\vspace*{.3in}


\maketitle 

\section{Introduction}

In recent years, advances in the field of electronics, telecommunication, computer and engineering sciences have led to a very strong increase, both in terms of speed and volume, in data being generated. Popular exponents of the present large-scale data era are companies as Facebook Inc. and Google Inc., of which the latter has already been processing more than 20 petabytes of data per day since 2008, but government organizations are also important players \citep{DEAN:2008,KALIL:2012}. 

Due to the declining costs of bandwidth, computing power and storage of data, it is expected that this trend will persist in the future. Inevitably, this necessitates the design of tools to gain insights into these large-scale data sets. Therefore, the design of data mining, statistical and machine learning algorithms to examine large amounts of data and support decision making is of key interest. Commonly used approaches in this research field are clustering, dimensionality reduction, filtering, classification and regression modeling. The focus of this paper is regression modeling, more specifically, semiparametric regression which represents a large class of regression models that allow for nonlinear effects in predictive models \citep{RUPPERT:2003}. The typical approach to address semiparametric regression modeling is by analyzing the data in one batch. This requires collecting all data before analysis and storing it on one machine. Having all data available at a central location is, however, unrealistic or not feasible for the large-scale setting. For example, Google Inc. designed and implemented a scalable distributed file system to meet its storage needs and, more recently, reported about Spanner, its scalable, multi-version, globally distributed, and synchronously-replicated database, having data centers spread all over the world \citep{GHEMAWAT:2003,CORBETT:2012}. Other examples of organizations that have their data distributed over multiple locations are federal departments and agencies, retail companies, but also peer-to-peer networks are part of this scenario. Moreover, in many contexts it can be in the interest of the different organizations (e.g. networks of retailers, hospitals) to combine their individual, potentially non-distributed, data sets to discover new knowledge for improved decision making. 

When large data sets are distributed over multiple machines or locations, moving the actual data is usually not a solution due to the associated communication complexity. Apart from the inefficiency, data confidentiality is another important reason to develop feasible alternatives, since different cooperating organizations may not be allowed or willing to share raw data. In this paper, the focus is on so-called horizontally distributed data, meaning that each data host stores different data subjects, but holds the same attributes \citep{DU:2004}. In the regression context this means that the different hosts store different samples and have all corresponding predictor variables available. The literature on regression modeling for horizontally distributed data sets is largely concerned with multivariate linear regression \citep{KARR:2005,KARR:2007,GHOSH:2012}. An exception to this is the study by \citet{GHOSH:2007} that presented an approach based on multivariate adaptive regression splines to incorporate more flexibility in the model. These approaches typically combine the output from local regression models (cf. ensemble learning) or combine local summary statistics to find the global regression model. Other notable studies are those by \citet{BOYD:2011} and \citet{RAM:2012} in the context of distributed convex optimization and \citet{PREDD:2006}. 

The focus of this paper, by contrast, is on semiparametric regression modeling for data sets that are horizontally partitioned over multiple hosts and the use of mean field variational Bayes (MFVB) for approximate inference \citep{WAINWRIGHT:2008,ORMEROD:2010}. Interpreting semiparametric regression in terms of graphical models offers an elegant and unified way to handle, for example, generalized additive models, geostatistical models, wavelet nonparametric regression models and their various combinations \citep{WAND:2011}. Moreover, MFVB provides a fast alternative to Markov chain Monte Carlo (MCMC) for fitting these models while it exhibits excellent accuracy for the models that this paper deals with. The methodology also enables handling of grouped data, within-subject correlation, automated regularization parameter inference and various (hierarchical) priors. Importantly, apart from point estimates, measures of uncertainty can be obtained in a straightforward way. 

While data sets have typically been processed in batch, nowadays, there is increasing interest in real-time systems that require so-called online stream-processing \citep{MICHALAK:2012}. Other studies outside of semiparametric regression that developed online methods for horizontally partitioned data are \citet{GUESTRIN:2004}, \citet{BHADURI:2008}, \citet{POZDNOUKHOV:2011} and \citet{YAN:2013}. Interestingly, the former three studies included mechanisms for concept drift in the algorithms for distributed regression. Online MFVB algorithms that make a single pass through the data have recently been developed. \citet{HOFFMAN:2010} and \citet{WANG:2011} proposed MFVB methods for latent Dirichlet allocation and the hierarchical Dirichlet process for topic modeling, respectively. \citet{TCHUMTCHOUA:2011} used online MFVB inference for high-dimensional correlated data and, very recently, \citet{LUTS:2013} proposed real-time semiparametric regression through MFVB approximate inference. However, it appears that these MFVB-based studies only dealt with non-distributed data sets for inference. Therefore, this paper demonstrates batch semiparametric regression for large-scale horizontally distributed data sets and real-time semiparametric regression for processing of horizontally distributed infinite data streams. The proposed algorithms provide exact solutions in the sense that an identical solution is obtained as when all data would have been available at a central location. In addition, this study proposes approaches for temporal adaptation for real-time semiparametric regression of distributed data streams, offering fully-automated regularization for evolving environments. A website is created for real-time demonstration of these methods on live airline data.

Section \ref{section2} provides background material on semiparametric regression and MFVB approximate inference. Semiparametric regression for distributed data sets is introduced in Section \ref{section3}. Both batch and real-time processing are treated. In Section \ref{section4} two approaches are presented to handle the issue of evolving environments for real-time semiparametric regression. Section \ref{section5} deals with two application areas of the proposed methodology: semiparametric regression in case of multiple data owners and within the MapReduce framework \citep{DEAN:2008,WHITE:2009}. A dynamic website that illustrates the methodology on live airline data is the focus of Section \ref{section6}. Closing remarks are made in Section \ref{section7}.

\section{Variational Bayesian inference for semiparametric regression} \label{section2}

Penalized splines are often used in the semiparametric regression literature to model nonlinear functional relationships \citep{RUPPERT:2009}. Consider the simple model

\begin{equation}
f(x_i)=\beta_0+\beta_1 x_i+\sum_{k=1}^K u_k z_k(x_i),\quad u_k\ \simind\ N(0,\sigma_u^2), \quad 1 \leq i \leq n,\\
\label{simpleSemipar1}
\end{equation}

\noindent with model parameters $\beta_0$, $\beta_1$, $u_1, \ldots, u_K$ and smoothing parameter $\sigma_u^2$, while $\simind$ denotes distributed independently. The $z_1(\cdot), \ldots, z_K(\cdot)$ represent spline basis functions and in this paper O'Sullivan splines, providing a close approximation to smoothing splines, are used for this purpose \citep{WAND:2008}. Note that (\ref{simpleSemipar1}) can be interpreted as a linear mixed model and leads to the following Bayesian Gaussian response model

\begin{equation}
\begin{array}{c}
y_i|\beta_0,\beta_1,u_1, \ldots , u_K \simind N\left(\beta_0+\beta_1 x_i+\sum_{k=1}^Ku_k z_k(x_i),\sigeps^2\right),\\
\null\\ 
1 \leq i \leq n,\quad u_k \simind N(0,\sigma_u^2),\quad \beta_0,\beta_1\simind N(0,\sigma_{\beta}^2),\\
\end{array}
\label{simpleSemipar2}
\end{equation}

\noindent for given smoothing parameter $\sigma_u^2$, error variance $\sigeps^2$ and positive hyperparameter $\sigma_{\beta}^2$. By assuming an arbitrary number of predictor variables and introducing uninformative priors for $\sigma_{u1}^2, \ldots, \sigma_{ur}^2$ and $\sigeps^2$ the more general representation of (\ref{simpleSemipar2}) becomes

\begin{equation}
\begin{array}{c}
\by|\,\bbeta,\bu,\sigeps^2 \sim 
N(\bX\bbeta+\bZ\bu,\sigeps^2\,\bI_n), \quad \bbeta\sim N(\bzero,\sigma_{\beta}^2\bI_p),\\
\null\\ 
\bu|\,\sigma_{u1}^2, \ldots,\sigma_{ur}^2\sim N(\bzero,\mbox{blockdiag}(\sigma_{u1}^2\,\bI_{K_1},\ldots,\sigma_{ur}^2\,\bI_{K_r})),\\
\null\\ 
\sigma_{u\ell}\simind\mbox{Half-Cauchy}(A_{u\ell}),\ 1\le\ell\le r,\quad 
\sigeps\sim\mbox{Half-Cauchy}(A_{\varepsilon}),\\
\end{array}
\label{lmm}
\end{equation}

\noindent where $\by$ is an $n\times1$ vector of response variables, $A_{\varepsilon}$ and $A_{u\ell}$ are positive hyperparameters, $\bbeta$ is a $p\times1$ vector of fixed effects, 
$\bu$ is a $(\sum^{r}_{l=1}K_l)\times1$ vector of random effects and $\bX$ and $\bZ$ corresponding design matrices. In this paper, all examples are based on the following values for the hyperparameters: $\sigma_{\beta}^2=10^8$ and $A_{\varepsilon}=A_{u\ell}=10^5$. Note that the variance parameters $\sigma_{u1}^2, \ldots, \sigma_{ur}^2$ correspond to the sub-blocks of $\bu$ having size $K_1,\ldots,K_r$, respectively. In this model the $\text{Half-Cauchy}(A)$ prior is such that the prior density of $\sigma$ is $p(\sigma)\propto\{1+(\sigma/A)^2\}^{-1},\ \sigma>0$. To obtain an equivalent, but more tractable model, the form $\sigma\sim\text{Half-Cauchy}(A)$ is replaced in (\ref{lmm}) by the auxiliary variable representation introduced in \citet{WAND:2011b}

\begin{equation}
\sigma^2|\,a\sim\text{Inverse-Gamma}\left(1/2,1/a\right),\quad a\sim\text{Inverse-Gamma}\left(1/2,1/A^2\right),\\ \nonumber
\end{equation}

\noindent where $v\sim\mbox{Inverse-Gamma}(A,B)$ if and only if its density function is

\begin{equation}
p(v)=B^A\Gamma(A)^{-1}\,v^{-A-1}\,\exp(-B/v),\quad v>0.\\ \nonumber
\end{equation}

\noindent As \citet{ZHAO:2006} reported, model (\ref{lmm}) is quite general and encompasses a large class of models, including simple random effects models, cross random effects models, nested random effects models, generalized additive models, semiparametric mixed models, bivariate smoothing and geoadditive models. For example, in the case of a simple semiparametric model with one predictor and a random intercept, (\ref{lmm}) reduces to

\begin{equation}
\begin{array}{c}
y_{ij}|\beta_0,\beta_1,u_1, \ldots , u_K,U_i,\sigeps^2\simind N(\beta_0+\beta_1\,x_{ij}+\sum_{k=1}^Ku_k z_k(x_{ij})+U_i,\sigeps^2),\\
\null\\
1\le i\le m, \quad 1\le j\le n_i, \quad
\beta_0,\beta_1\simind N(0,\sigma_{\bbeta}^2),\quad u_k \simind N(0,\sigma_u^2),\\
\null\\
U_i|\,\sigma_U^2\simind N(0,\sigma_U^2), \quad \sigma_u\sim\mbox{Half-Cauchy}(A_u),\\
\null\\
\sigma_U\sim\mbox{Half-Cauchy}(A_U), \quad \sigeps\sim\mbox{Half-Cauchy}(A_{\varepsilon}),\\
\vspace{-1.6cm}
\end{array}
\vspace{1.6cm}
\label{eq:randInt}
\end{equation}

\noindent where $(x_{ij},y_{ij})$ represents the $j$th predictor/response pair for the $i$th group, with $n_{i}$ denoting the number of subjects in group $i$ and $m$ the total number of groups. An extension of random intercept model (\ref{eq:randInt}) is used for the real-life example in Section \ref{section6}.
  
MFVB is a class of methods relying on approximate inference of posterior density functions \citep{WAINWRIGHT:2008,ORMEROD:2010}. A mean field approximation is founded upon approximating the posterior density function $p(\btheta|\by)$, e.g. parameter vector $\btheta = [\bbeta,\bu,a_{u1},\ldots,$ $a_{ur},a_{\varepsilon},
\sigma_{u1}^2,\ldots,\sigma_{ur}^2,$ $\sigeps^2]^{T}$ for model (\ref{lmm}), by a product form $q(\btheta) = \prod_{i=1}^{d} q_i(\btheta_i)$. The choice of the $q_i(\btheta_i)$ density functions is guided by the notion of Kullback-Leibler divergence

\begin{equation}
\int q(\btheta) \log\left\{\frac{q(\btheta)} {p(\btheta|\by)}\right\}\,d\btheta,\\ 
\label{KL}
\end{equation}

\noindent such that the distance between $\prod_{i=1}^{d} q_i(\btheta_i)$ and $p(\btheta|\by)$ is minimized. It can be shown that an equivalent optimization problem corresponds to maximizing the so-called lower bound on the marginal likelihood $p(\by)$,

\begin{equation}
\punder(\by;q)\equiv\exp \left[ \int q(\btheta) \log\left\{\frac{p(\btheta,\by)} {q(\btheta)}\right\}\,d\btheta\right].\\ \nonumber
\end{equation}

\noindent The optimal $q^*_i(\btheta_i)$ density functions, in terms of minimizing the Kullback-Leibler divergence in (\ref{KL}), are known to satisfy

\begin{equation}
q^*_i(\btheta_i) \propto \exp \left[ \mathop{\mathlarger{\mathlarger{\mathlarger{\int}}}} {\left\{ \prod_{j\neq i} q_{j}(\btheta_j) \right\} \log p(\btheta,\by) \, d\btheta_{-i}} \right]\nonumber,\\
\end{equation}

\noindent where $\btheta_{-i} = [\btheta_{1},\ldots,\btheta_{i-1},\btheta_{i+1},\ldots,\btheta_{d}]^T$.

Although MFVB is limited in its approximation accuracy when compared to MCMC, which can be made arbitrarily accurate by increasing the Monte Carlo sample sizes, the latter is much slower than MFVB. Moreover, the accuracy of MFVB for the models that are considered in this paper is typically excellent.

For the mixed model in (\ref{lmm}) the mean field approximation and chosen product form 

\begin{equation}
\begin{array}{l}
p(\bbeta,\bu,a_{u1},\ldots,a_{ur},a_{\varepsilon},\sigma_{u1}^2,\ldots,\sigma_{ur}^2,\sigeps^2|\by) \\
\null\\
\hspace{3cm} \approx q(\bbeta,\bu,a_{u1},\ldots,a_{ur},a_{\varepsilon},\sigma_{u1}^2,\ldots,\sigma_{ur}^2,\sigeps^2)\\
\null\\
\hspace{3cm} \approx q(\bbeta,\bu,a_{u1},\ldots,a_{ur},a_{\varepsilon}) \,q(\sigma_{u1}^2,\ldots,\sigma_{ur}^2,\sigeps^2),\nonumber \\
\end{array}
\end{equation}

\noindent lead to the following optimal product density functions: $q^*(\bbeta,\bu,a_{u1},\ldots,a_{ur},a_{\varepsilon})$
is the product of the $ 
N(\bmu_{q(\bbeta,\bu)},\bSigma_{q(\bbeta,\bu)})$ density function,
Inverse-Gamma$(1,$ $B_{q(a_{u\ell})})$ density functions, $1\le\ell\le r$, and the Inverse-Gamma$(1,B_{q(a_{\varepsilon})})$ density function, while
$q^*(\sigma_{u1}^2,\ldots,\sigma_{ur}^2,\sigeps^2)$ is the product of
Inverse-Gamma$(\smhalf(K_{\ell}+1),B_{q(\sigma^2_{u\ell})})$
density functions for $1\le\ell\le r$ and the Inverse-Gamma$(\smhalf(n+1),B_{q(\sigeps^2)})$ density function. Notice that this solution results in so-called induced factorizations. For example, the factorization $q(\bbeta,\bu,a_{u1},\ldots,a_{ur},a_{\varepsilon})=q(\bbeta,\bu)\,q(a_{u1}),\ldots,q(a_{ur})\,q(a_{\varepsilon})$ is not assumed a priori. 

Since the optimal parameters in the $q^*$ density functions are interrelated, for example,

\begin{equation}
\bSigma_{q(\bbeta,\bu)}= \left[\mu_{q(1/\sigeps^2)}\,\bC^T\bC+\mbox{blockdiag}\{\sigma_{\beta}^{-2}\,\bI_p,\mu_{q(1/\sigma_{u1}^2)}\bI_{K_1},\ldots, \mu_{q(1/\sigma_{ur}^2)}\bI_{K_r}\} \right]^{-1}, \nonumber\\
\end{equation}

\noindent with $\bC = [\bX \, \bZ]$, the iterative coordinate ascent Algorithm \ref{algNonDistBatch} is used to compute the optimal densities where the logarithm of the lower bound equals
  
\begin{eqnarray*}
\log\,\punder(\by;q)&=& \frac{p+\sum_{\ell=1}^{r} K_\ell}{2} - \frac{n}{2} \log(2\pi) - (r+1) \log(\pi) - \frac{p}{2} \log(\sigma_{\beta}^{2}) + \frac{1}{2} \log(|\bSigma_{q(\bbeta,\bu)}|) \\[0.75ex]
&& + \log\left(\Gamma\left(\frac{n+1}{2}\right)\right) - \frac{1}{2\sigma_{\beta}^2} \{ ||\bmu_{q(\bbeta)}||^2 +\mbox{tr}(\bSigma_{q(\bbeta)}) \} - \left(\frac{n+1}{2} \right) \log(B_{q(\sigeps^2)})  \\[0.75ex]
&& + \mu_{q(1/\aeps)} \mu_{q(1/\sigeps^2)} - \log(\Aeps) - \log(B_{q(a_{\varepsilon})}) + \sum_{\ell=1}^{r} \Bigg\{ \log\left(\Gamma\left(\frac{K_\ell+1}{2}\right)\right) - \log(A_{u\ell}) \\[0.75ex]
&& - \log(B_{q(a_{u\ell})}) - \left(\frac{K_\ell+1}{2}\right) \log(B_{q(\sigma^2_{u\ell})}) + \mu_{q(1/a_{u\ell})} \mu_{q(1/\sigma^2_{u\ell})} \Bigg\}.\\ \nonumber
\end{eqnarray*}

\begin{algorithm}
\noindent \caption{\textit{Mean field variational Bayes algorithm for obtaining the parameters in the optimal densities for the Gaussian linear mixed model (\ref{lmm}).}}
\label{algNonDistBatch}
\begin{algorithmic}[1]
\REQUIRE $\bC, \by, n, p, K_{\ell}, \mu_{q(1/\sigeps^2)}, \Aeps, \mu_{q(1/\sigma_{u\ell}^2)}, A_{u\ell}, \sigma_{\beta}^{2} \, \text{with} \, 1\le\ell\le r$
\WHILE{the increase in $\log \punder(\by;q)$ is significant}
    \STATE $\bSigma_{q(\bbeta,\bu)}\leftarrow \left[\mu_{q(1/\sigeps^2)}\,\bC^T\bC+ \mbox{blockdiag}\{\sigma_{\beta}^{-2}\,\bI_p, \mu_{q(1/\sigma_{u1}^2)}\bI_{K_1},\ldots, \mu_{q(1/\sigma_{ur}^2)}\bI_{K_r}\} \right]^{-1}$    
   	\STATE $\bmu_{q(\bbeta,\bu)}\leftarrow 
\mu_{q(1/\sigeps^2)}\,\bSigma_{q(\bbeta,\bu)}\,\bC^T\by;\quad \mu_{q(1/\aeps)}\leftarrow 1/\{\mu_{q(1/\sigeps^2)}+\Aeps^{-2}\}$
		\STATE $\mu_{q(1/\sigeps^2)}\leftarrow 
\displaystyle{\frac{n+1}{2\,\mu_{q(1/\aeps)}
+\by^T\by-2\bmu_{q(\bbeta,\bu)}^T\bC^T\by
+\mbox{tr}[(\bC^T\bC)\{\bSigma_{q(\bbeta,\bu)}+\bmu_{q(\bbeta,\bu)}
\bmu_{q(\bbeta,\bu)}^T\}]}}$
		\FOR{$\ell = 1 \to r$}
				\STATE $\mu_{q(1/a_{u\ell})}\leftarrow 1/\{\mu_{q(1/\sigma_{u\ell}^2)}+A_{u\ell}^{-2}\};\quad \mu_{q(1/\sigma_{u\ell}^2)}\leftarrow 
\displaystyle{\frac{K_{\ell}+1}{2\,\mu_{q(1/\auell)}
+\Vert\bmu_{q(\bu_{\ell})}\Vert^2
+\mbox{tr}(\bSigma_{q(\bu_{\ell})})}}$
		\ENDFOR
\ENDWHILE\end{algorithmic}
\end{algorithm}

\section{Semiparametric regression for distributed data sets} \label{section3}

Specifying an appropriate regression model might be difficult when data are spread over multiple hosts and there is no opportunity to inspect the combined data set. In these circumstances, semiparametric regression represents a viable alternative to multivariate linear regression, as the latter requires having to choose in advance which polynomial terms to include or transformations to apply. On the other hand, the MFVB approach that was presented in Section \ref{section2} includes inference for the smoothing parameters $\sigma^2_{u\ell}, \,1\le\ell\le r$ and, as a consequence, offers fully-automated fitting of flexible relationships between predictors and the dependent variable. The following sections explain how to perform semiparametric regression for distributed data in the batch and the real-time setting.

\subsection{Batch processing}

Algorithm \ref{algNonDistBatch} relies on having the data for all $n$ samples in one location and receives these as input via $\bC$ and $\by$. The crucial piece that allows extending Algorithm \ref{algNonDistBatch} towards the distributed setting is how it uses the data: it only depends on the data through the quantities $\bC^T\bC$, $\bC^T\by$, $\by^T\by$ and $n$. Assuming that there are $h$ different locations that host data, i.e. $\bC_g$, $\by_g$ and $n_g$, $1\leq g \leq h$, the following straightforward relationships can be used: $\bC^T\bC = \sum_{g=1}^{h}\bC_g^T\bC_g$, $\bC^T\by = \sum_{g=1}^{h}\bC_g^T\by_g$, $\by^T\by = \sum_{g=1}^{h}\by_g^T\by_g$ and $n = \sum_{g=1}^{h}n_g$.

Algorithm \ref{algDistBatch} summarizes the procedure for batch semiparametric regression for distributed data sets. Note that $P=p+\sum_{l=1}^{r}K_l$ denotes the number of columns of $\bC$. Each host performs the computation of the summary statistics locally and Algorithm \ref{algDistBatch} only relies on those summaries. Therefore, there is no need to send the actual raw data over the network, thereby saving bandwidth and speeding up the algorithm. This approach is particularly useful for large-scale data sets having large sample sizes and it avoids security risks by data being flooded through the network. In addition, all hosts can generate the local summary statistics simultaneously, but Algorithm \ref{algDistBatch} can only start from the moment that all hosts have finished their local computations. The total number of parameters that each host has to send equals $P(P+1)/2+P+2$ since $\bC^T\bC$ is symmetric. Depending on the structure of $\bC$, this number can further be reduced. Section \ref{section3:b} illustrates this for the random intercept model (\ref{eq:randInt}). Note that Algorithm \ref{algDistBatch} assumes the existence of another party, called \textit{combiner} in this paper, which receives the local summary statistics from the data hosts and manages the global semiparametric regression. However, as Section \ref{section5:a} points out, the existence of a separate party is in fact not a requirement.

A potential issue with the proposed method is that the spline basis functions have to be set without having the combined data set available. For example, a set of knot positions may need to be specified. For many applications it is simple to specify the range of possible values beforehand. For example, for a predictor variable corresponding to outside temperature in degrees Celsius or wind speed in knots, equidistantly positioning knots within a reasonable range represents an effective approach. Dealing with grouped data, e.g. within the context of random intercept model (\ref{eq:randInt}), involves similar issues. Algorithm \ref{algDistBatch} requires specifying the number and kind of groups a priori. Again, for many applications this is not a problem. For example, when flights are grouped per airline, the total number of possible airlines can be determined beforehand. If these assumptions are not reasonable, some adjustments have to be made.  

\vspace{0.3cm}
\begin{algorithm}
\noindent \caption{\textit{Batch mean field variational Bayes algorithm for obtaining the parameters in the optimal densities for the Gaussian linear mixed model (\ref{lmm}) in case of distributed data sets.}}
\label{algDistBatch}
\begin{algorithmic}[1]
\REQUIRE $p, P, K_{\ell}, \mu_{q(1/\sigeps^2)}, \Aeps, \mu_{q(1/\sigma_{u\ell}^2)}, A_{u\ell}, \sigma_{\beta}^{2} \, \text{with} \, 1\le\ell\le r$
\STATE $\bC^T\bC \leftarrow \bzero_{P \times P}; \quad \bC^T\by \leftarrow \bzero_{P \times 1}; \quad \by^T\by \leftarrow 0; \quad n \leftarrow 0$
		\FOR{$g = 1 \to h$}
				\STATE $\text{retrieve} \, \, \bC_g^T\bC_g, \, \, \bC_g^T\by_g, \, \, \by_g^T\by_g \, \, \text{and} \, \, n_g \, \, \text{from host} \, g$
				\STATE $\bC^T\bC \leftarrow \bC^T\bC + \bC_g^T\bC_g; \quad \bC^T\by \leftarrow \bC^T\by + \bC_g^T\by_g ; \quad \by^T\by \leftarrow \by^T\by + \by_g^T\by_g; \quad n \leftarrow n + n_g$
		\ENDFOR
\WHILE{the increase in $\log \punder(\by;q)$ is significant}
    \STATE $\bSigma_{q(\bbeta,\bu)}\leftarrow \left[\mu_{q(1/\sigeps^2)}\,\bC^T\bC+ \mbox{blockdiag}\{\sigma_{\beta}^{-2}\,\bI_p, \mu_{q(1/\sigma_{u1}^2)}\bI_{K_1},\ldots, \mu_{q(1/\sigma_{ur}^2)}\bI_{K_r}\} \right]^{-1}$    
   	\STATE $\bmu_{q(\bbeta,\bu)}\leftarrow 
\mu_{q(1/\sigeps^2)}\,\bSigma_{q(\bbeta,\bu)}\,\bC^T\by;\quad \mu_{q(1/\aeps)}\leftarrow 1/\{\mu_{q(1/\sigeps^2)}+\Aeps^{-2}\}$
		\STATE $\mu_{q(1/\sigeps^2)}\leftarrow 
\displaystyle{\frac{n+1}{2\,\mu_{q(1/\aeps)}
+\by^T\by-2\bmu_{q(\bbeta,\bu)}^T\bC^T\by
+\mbox{tr}[(\bC^T\bC)\{\bSigma_{q(\bbeta,\bu)}+\bmu_{q(\bbeta,\bu)}
\bmu_{q(\bbeta,\bu)}^T\}]}}$
		\FOR{$\ell = 1 \to r$}
				\STATE $\mu_{q(1/a_{u\ell})}\leftarrow 1/\{\mu_{q(1/\sigma_{u\ell}^2)}+A_{u\ell}^{-2}\};\quad \mu_{q(1/\sigma_{u\ell}^2)}\leftarrow 
\displaystyle{\frac{K_{\ell}+1}{2\,\mu_{q(1/\auell)}
+\Vert\bmu_{q(\bu_{\ell})}\Vert^2
+\mbox{tr}(\bSigma_{q(\bu_{\ell})})}}$
		\ENDFOR
\ENDWHILE
\end{algorithmic}
\end{algorithm}

\subsubsection{Illustration for Sydney property rental data} \label{section3:a2}

This section illustrates distributed batch semiparametric regression by analyzing data from the residential property rental market in Sydney, Australia. With more than a thousand real estate offices, the Sydney real estate market is a highly competitive one. All together, 1447 real estate offices hosted data belonging to 150471 properties during the period 9th May, 2012 and 25th May, 2013. In this example, these data are processed by Algorithm \ref{algDistBatch} using the model
\begin{equation}
\begin{array}{l}
\log(\mbox{\texttt{weekly\,\,rent}}_{i})|\,\bbeta,
\bu_2,\bu_3,\bu_4,\bu_5,
\sigeps^2\simind\\[1ex] 
\qquad
N(\beta_0 + \beta_1\,\mbox{\texttt{house}}_{i} +
f_2(\mbox{\texttt{number\,\,of\,\,bedrooms}}_{i})\\[1ex]
\qquad + f_3(\mbox{\texttt{number\,\,of\,\,bathrooms}}_{i} )
+f_4(\mbox{\texttt{number\,\,of\,\,car\,\,spaces}}_{i} )\\[1ex] 
\qquad+ f_5(\mbox{\texttt{longitude}}_{i},\mbox{\texttt{latitude}}_{i}),\sigeps^2),
\end{array}
\label{eq:SydneyRealEstate}
\end{equation}
where $\mbox{\texttt{weekly rent}}_{i}$ is the weekly rental amount in Australian dollars of the $i$th property, $\mbox{\texttt{house}}_{i}$ is an indicator of the $i$th property being a house, townhouse or villa versus an apartment, and $\mbox{\texttt{number\,\,of\,\,bedrooms}}_{i}$ is the number of bedrooms in the $i$th property. Variables concerning the number of bathrooms and car spaces are defined in a similar way. The geographical location of the $i$th property
is included by the variables $\texttt{longitude}_{i}$ and $\texttt{latitude}_{i}$. To execute Algorithm \ref{algDistBatch}, the fixed effect regression coefficients $\beta_0$, $\beta_1$  and the linear contributions to $f_2,\ldots,f_5$ are stored in $\bbeta$, while the spline basis coefficients for $f_2,\ldots,f_5$ are stored in $\bu_2,\ldots,\bu_5$. The estimate of $f_5$ is based
on bivariate thin plate splines \citep{RUPPERT:2003}.

\begin{figure}[!ht]
\centering
{\includegraphics[width=\textwidth]{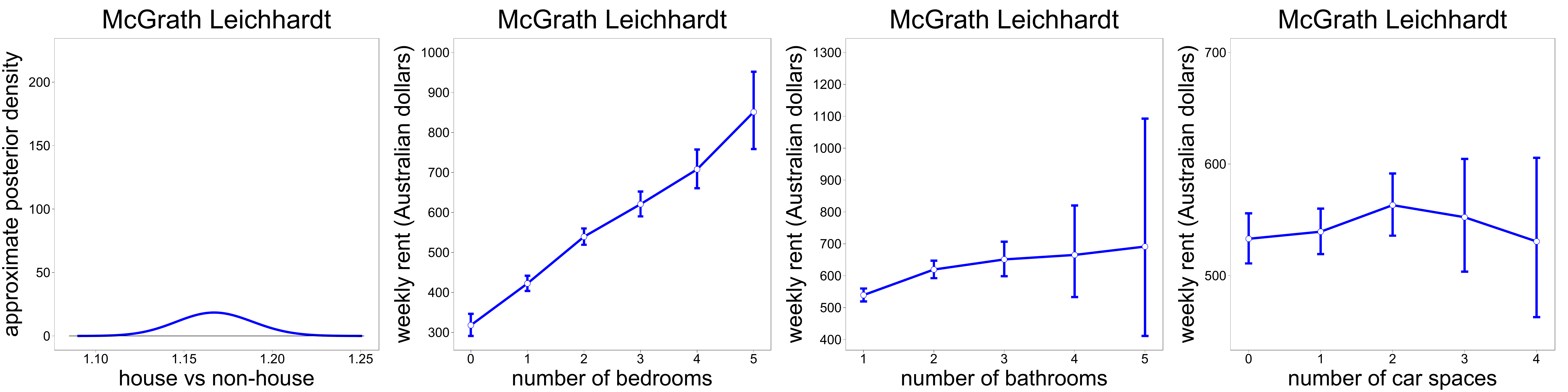}}\\
\textbf{}
{\includegraphics[width=\textwidth]{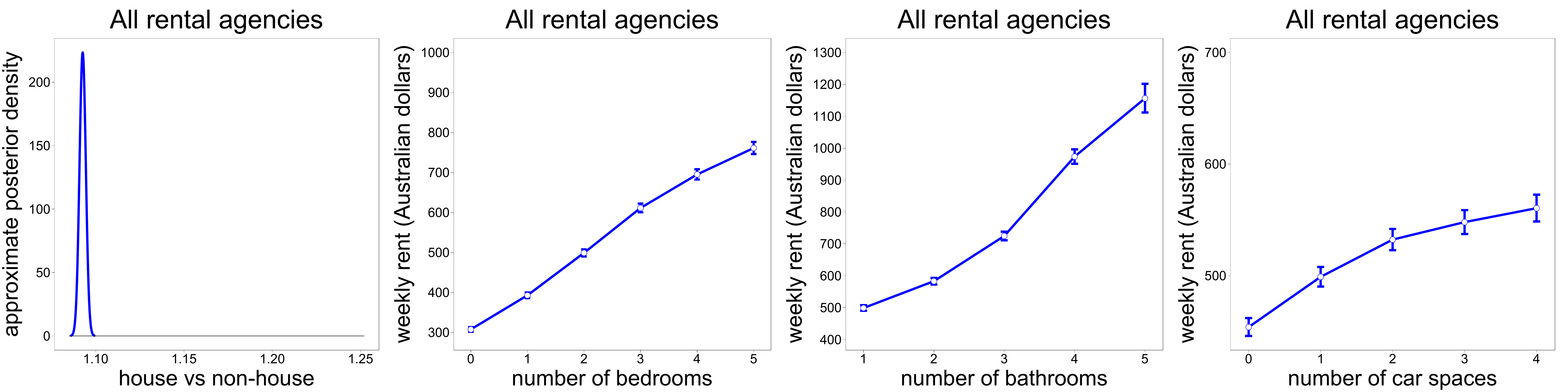}}
\caption{\it Approximate posterior density functions, regression fits and corresponding 95\% credible sets for the Sydney property rental data example in Section \ref{section3:a2}. The first column displays the impact of the property being a house or not, while the other three columns visualize the effects of the number of bedrooms, bathrooms and car spaces on the weekly rent for apartments. The top row results are based on data from the real estate agency \texttt{McGrath Leichhardt} only, whereas the bottom row displays results based on data hosted by 1447 real estate offices.}
\label{fig:batch1} 
\end{figure}

\begin{figure}[!ht]
\centering
{\includegraphics[trim = 10mm 8mm 25mm 20mm, clip, height=9.1cm]{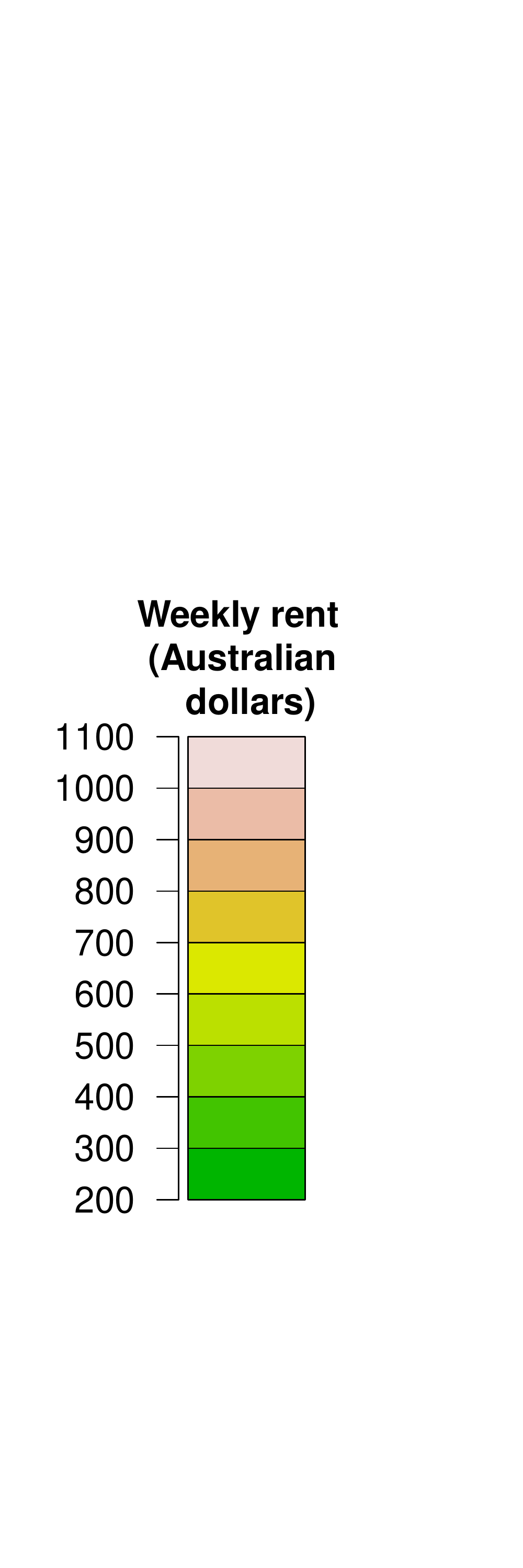}\includegraphics[width=9cm]{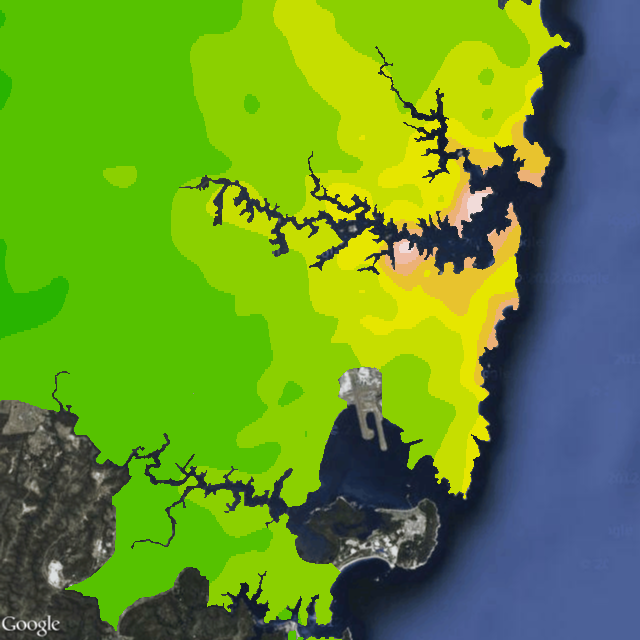}}
\caption{\it Color-coded geographical map of Sydney displaying the estimated mean weekly rent for a two bedroom apartment with
one bathroom and one car space for various geographical locations based on data hosted by 1447 real estate offices as explained in Section \ref{section3:a2}.}
\label{fig:batch2} 
\end{figure}

Figure \ref{fig:batch1} shows various regression summaries resulting from fitting of (\ref{eq:SydneyRealEstate}) using Algorithm \ref{algNonDistBatch} on data from real estate agency \texttt{McGrath Leichhardt} only, corresponding to 436 properties, and the combined result from Algorithm \ref{algDistBatch} based on data hosted by 1447 real estate offices (i.e. 150471 properties). As expected, the estimates based on combining information from multiple hosts are more reliable and the figure shows that more narrow 95\% credible sets are obtained. The approximate posterior density function for $\beta_1$ shows that the average rental amount for houses is 9.5\% higher than for apartments after correcting for all other covariates. The remaining panels show the increase in rental amount when the property includes more bedrooms, bathrooms or car spaces. Finally, a color-coded geographical map of Sydney, based on the data from 1447 hosts, displays the weekly rent for a two bedroom apartment with one bathroom and one car space for various geographical locations (Figure \ref{fig:batch2}).

\subsection{Real-time processing} \label{section3:b}

The implementation of online regression methods in real-time systems supports incremental calculations of regression results when new data arrive \citep{LUTS:2013}. This section focuses on the more complicated setting where $h$ different hosts simultaneously receive different streams of data, independent from each other. While the hosts can individually monitor their data streams and perform semiparametric regression in real-time, the real-time combined regression result based on all $h$ distributed data streams is of primary interest in this paper. Algorithm \ref{algDistOnline} summarizes how the combiner manages the global semiparametric regression in an incremental manner for such a setting.

First, the individual data hosts process their data streams locally in real-time. This includes repeatedly extracting the dependent variable and the predictor variables from the stream, but may involve additional preprocessing. Online semiparametric regression at each host is performed according to Algorithm 3 of \citet{LUTS:2013}. In addition to this, each host stores the summary statistics for its incoming data $\bc_{\text{new}}\,\bc_\text{new}^T$, $\bc_\text{new}\, y_{\text{new}}$ and  $y^2_{\text{new}}$ in a local buffer. Once the local buffer of a host exceeds a threshold size, the sum of the buffer's local summary statistics, i.e. $\bC_b^T\bC_b$, $\bC_b^T\by_b$, $\by_b^T\by_b$ and $n_b$, is sent to the combiner and the buffer is emptied. A buffer at the site of the data host enables it to regulate data traffic and, depending on its size and the rate at which data are coming in, the combiner will receive the local summary statistics with a certain delay. The combiner has its own buffer where the summaries $\bC_b^T\bC_b$, $\bC_b^T\by_b$, $\by_b^T\by_b$ and $n_b$ from the different hosts are stored. The use of a buffer at the combiner site allows a difference between the rate at which summary statistics are received from the data hosts and the rate at which they can be processed by the combiner via Algorithm \ref{algDistOnline}. This setup enables the data hosts to operate independently and asynchronously from each other. In addition, they can simultaneously process the different streams, speeding up the computation of the combined semiparametric regression result.

\begin{algorithm}
\noindent \caption{\textit{Online mean field variational Bayes algorithm for obtaining the parameters in the optimal densities for the Gaussian linear mixed model (\ref{lmm}) in case of distributed data sets.}}
\label{algDistOnline}
\begin{algorithmic}[1]
\REQUIRE $p, P, K_{\ell}, \mu_{q(1/\sigeps^2)}, \Aeps, \mu_{q(1/\sigma_{u\ell}^2)}, A_{u\ell}, \sigma_{\beta}^{2} \, \text{with} \, 1\le\ell\le r$
\STATE $\bC^T\bC \leftarrow \bzero_{P \times P}; \quad \bC^T\by \leftarrow \bzero_{P \times 1}; \quad \by^T\by \leftarrow 0; \quad n \leftarrow 0$
\WHILE{new data available in buffer}
		\STATE $B \leftarrow \text{number of items in buffer}$
		\STATE $\text{read items from buffer and compute} \, \, \sum_{b=1}^{B}\bC_b^T\bC_b, \, \, \sum_{b=1}^{B} \bC_b^T\by_b, \, \, \sum_{b=1}^{B} \by_b^T\by_b \, \, \text{and} \, \, \sum_{b=1}^{B} n_b$
		\STATE $\bC^T\bC \leftarrow \bC^T\bC + \sum_{b=1}^{B}\bC_b^T\bC_b; \quad \bC^T\by \leftarrow \bC^T\by + \sum_{b=1}^{B} \bC_b^T\by_b$
		\STATE $\by^T\by \leftarrow \by^T\by + \sum_{b=1}^{B} \by_b^T\by_b; \quad n \leftarrow n + \sum_{b=1}^{B} n_b$
		\STATE $\bSigma_{q(\bbeta,\bu)}\leftarrow \left[\mu_{q(1/\sigeps^2)}\,\bC^T\bC+ \mbox{blockdiag}\{\sigma_{\beta}^{-2}\,\bI_p, \mu_{q(1/\sigma_{u1}^2)}\bI_{K_1},\ldots, \mu_{q(1/\sigma_{ur}^2)}\bI_{K_r}\} \right]^{-1}$    
   	\STATE $\bmu_{q(\bbeta,\bu)}\leftarrow 
\mu_{q(1/\sigeps^2)}\,\bSigma_{q(\bbeta,\bu)}\,\bC^T\by;\quad \mu_{q(1/\aeps)}\leftarrow 1/\{\mu_{q(1/\sigeps^2)}+\Aeps^{-2}\}$
		\STATE $\mu_{q(1/\sigeps^2)}\leftarrow 
\displaystyle{\frac{n+1}{2\,\mu_{q(1/\aeps)}
+\by^T\by-2\bmu_{q(\bbeta,\bu)}^T\bC^T\by
+\mbox{tr}[(\bC^T\bC)\{\bSigma_{q(\bbeta,\bu)}+\bmu_{q(\bbeta,\bu)}
\bmu_{q(\bbeta,\bu)}^T\}]}}$
		\FOR{$\ell = 1 \to r$}
				\STATE $\mu_{q(1/a_{u\ell})}\leftarrow 1/\{\mu_{q(1/\sigma_{u\ell}^2)}+A_{u\ell}^{-2}\};\quad \mu_{q(1/\sigma_{u\ell}^2)}\leftarrow 
\displaystyle{\frac{K_{\ell}+1}{2\,\mu_{q(1/\auell)}
+\Vert\bmu_{q(\bu_{\ell})}\Vert^2
+\mbox{tr}(\bSigma_{q(\bu_{\ell})})}}$
		\ENDFOR
\ENDWHILE
\end{algorithmic}
\end{algorithm}

Note that Algorithm \ref{algDistOnline} initializes the summary statistics in line 1 to zero and requires starting values for $\mu_{q(1/\sigeps^2)}$ and $\mu_{q(1/\sigma_{u\ell}^2)}$. Section 2.1.1 of \citet{LUTS:2013} explains that good initialization by means of a so-called warm-up step can be important for convergence of the real-time semiparametric regression approach. For clarity of presentation, this warm-up step was not included in Algorithm \ref{algDistOnline}. Although experiments have shown that warming-up is in the first place important for wavelet regression and logistic regression (cf. \citet{LUTS:2013}), it can also easily be incorporated in Algorithm \ref{algDistOnline}. All it requires is running batch Algorithm \ref{algNonDistBatch} on a subset of data and using the summary statistics and obtained estimates as starting values for Algorithm \ref{algDistOnline}.

Closer inspection of Algorithm \ref{algDistOnline} reveals that line 6 involves inverting a matrix of size $P \times P$, with $P = p+\sum_{\ell=1}^{r}K_\ell$. As also noted by \citet{SMITH:2008} in the context of frequentist inference for additive mixed models, na{\"i}ve implementation of line 6 can be extremely inefficient for grouped data as in (\ref{eq:randInt}). Moreover, since Algorithm \ref{algDistOnline} aims to run in an online fashion on large-scale data with potentially many groups, and as a consequence large $P$, it is important to optimize this line of code. \citet{SMITH:2008} outline a procedure for which the variance calculations are linear in the number of groups, but omit the computation of correlations between any two groups. Algorithm \ref{algDistOnline}, however, does require calculating these inter-group correlations since the full matrix $\bSigma_{q(\bbeta,\bu)}$ is needed to compute $\bmu_{q(\bbeta,\bu)}$, for example. The following paragraphs explain how line 6 can be solved in more efficient way for grouped data as for example the live example in Section \ref{section6}.

Assume that $\bC = [\bX \, \bZ_1 \, \bZ_2]$, where the original design matrix $\bZ$ is divided into a design matrix that is only related to the $K_r$ random intercepts, i.e. $\bZ_2$, and a design matrix for all the rest, i.e. $\bZ_1$, including spline basis functions. This enables the block decomposition 

\begin{equation}
\renewcommand{\arraystretch}{1.7}
\begin{array}{rl}
\bM \ \equiv \ \bSigma_{q(\bbeta,\bu)}^{-1} & = \ \mu_{q(1/\sigeps^2)}
\left[ \begin{array}{cc|c}
       \bX^T \bX + \mu_{q(1/\sigeps^2)}^{-1} \bG_1 & \bX^T \bZ_1 & \bX^T \bZ_2 \\
       \bZ^T_1 \bX & \bZ_1^T \bZ_1 + \mu_{q(1/\sigeps^2)}^{-1} \bG_2 & \bZ_1^T \bZ_2 \\ \hline 
       \bZ_2^T \bX & \bZ_2^T \bZ_1 & \bZ_2^T \bZ_2 + \mu_{q(1/\sigeps^2)}^{-1} \bG_3
\end{array} \right]\\
\null\\
& = \ \mu_{q(1/\sigeps^2)}
\renewcommand{\arraystretch}{1}
\left[ \begin{array}{cc}
\bM_{11} & \bM_{12}\\[0.3em]
\bM_{21} & \bM_{22}\\[0.3em]
\end{array} \right], \nonumber
\end{array}
\end{equation}

\noindent where $\bG_1 = \sigma_{\beta}^{-2}\,\bI_p$, $\bG_2 = \mbox{blockdiag}\{\mu_{q(1/\sigma_{u1}^2)}\bI_{K_1},\ldots, \mu_{q(1/\sigma_{ur-1}^2)}\bI_{K_{r-1}}\}$ and $\bG_3 = \mu_{q(1/\sigma_{ur}^2)}$ $\bI_{K_r}$. The rules for computing the inverse of a block-partitioned matrix give

\begin{equation}
\bM^{-1} \ \equiv \ \bSigma_{q(\bbeta,\bu)} = \mu_{q(1/\sigeps^2)}^{-1} \left[ \begin{array}{cc}
\bM^{11} & \bM^{12}\\[0.3em]
\bM^{21} & \bM^{22}\\[0.3em]
\end{array} \right], 
\label{inverse}
\end{equation}

\noindent with $\bM^{11}=(\bM_{11}-\bM_{12}\bM_{22}^{-1}\bM_{21})^{-1}$, $\bM^{12}=-\bM^{11}\bM_{12}\bM_{22}^{-1}$, $\bM^{21}=(\bM^{12})^T$ and $\bM^{22}=\bM_{22}^{-1}+\bM_{22}^{-1}\bM_{21}\bM^{11}\bM_{12}\bM_{22}^{-1}$ \citep{HARVILLE:2000}. Dealing with a large number of groups results in the relationship $K_r \gg p + \sum_{\ell=1}^{r-1} K_l$. In these circumstances, the straightforward matrix multiplications $\bX^T \bX$, $\bX^T \bZ_1$ and $\bZ_1^T \bZ_1 $ are relatively inexpensive. As also explained in \citet{SMITH:2008}, $\bZ_2$ has a special structure because of the random intercept design, thereby making the computation of $\bX^T \bZ_2$ and $\bZ_1^T \bZ_2$ efficient. The biggest inverse that is needed in (\ref{inverse}) is $\bM_{22}^{-1}$, but since $\bM_{22}$ is diagonal it can be obtained in $K_r$ steps. The final step to obtain $ \bSigma_{q(\bbeta,\bu)}^{-1} $ is computing $\bM^{22}$. Whereas \citet{SMITH:2008} only compute the diagonal elements of this matrix, Algorithm \ref{algDistOnline} requires all unique entries of this symmetric matrix. Denoting $\bM_{12}=[\bh_1 , \ldots , \bh_{K_r}]$, the elements of $\bM^{22}$ are

\begin{equation}
\begin{array}{rl}
\bM^{22}_{ii} & = \frac{\mu_{q(1/\sigeps^2)}}{n_i \mu_{q(1/\sigeps^2)}+\mu_{q(1/\sigma_{ur}^2)}} \left(1+ \frac{\mu_{q(1/\sigeps^2)} \, \bh_i^T\bM^{11}\bh_i}{n_i \mu_{q(1/\sigeps^2)}+\mu_{q(1/\sigma_{ur}^2)}} \right),\\
\null\\
\bM^{22}_{ij} & = \frac{\mu^2_{q(1/\sigeps^2)} \, \bh_i^T\bM^{11}\bh_j}{\left(n_i \mu_{q(1/\sigeps^2)}+\mu_{q(1/\sigma_{ur}^2)}\right)\left(n_j \mu_{q(1/\sigeps^2)}+\mu_{q(1/\sigma_{ur}^2)}\right)}, 
\quad i\neq j,\\ \nonumber
\end{array}
\end{equation}

\noindent with $n_i$ the number of subjects in group $i$. Observe that the $K_r(K_r + 1)/2$ unique entries of $\bM^{22}$ can be computed in parallel. In addition, grouped data sets enable a further, significant reduction in unique entries to be transferred from host to combiner as $\bZ^T_2 \bZ_2$ is diagonal.

\subsubsection{Illustration for simulated data} \label{section3:b2}

\begin{figure}[!ht]
\centering
{\includegraphics[trim = 23mm 0mm 0mm 20mm, clip, width=\textwidth]{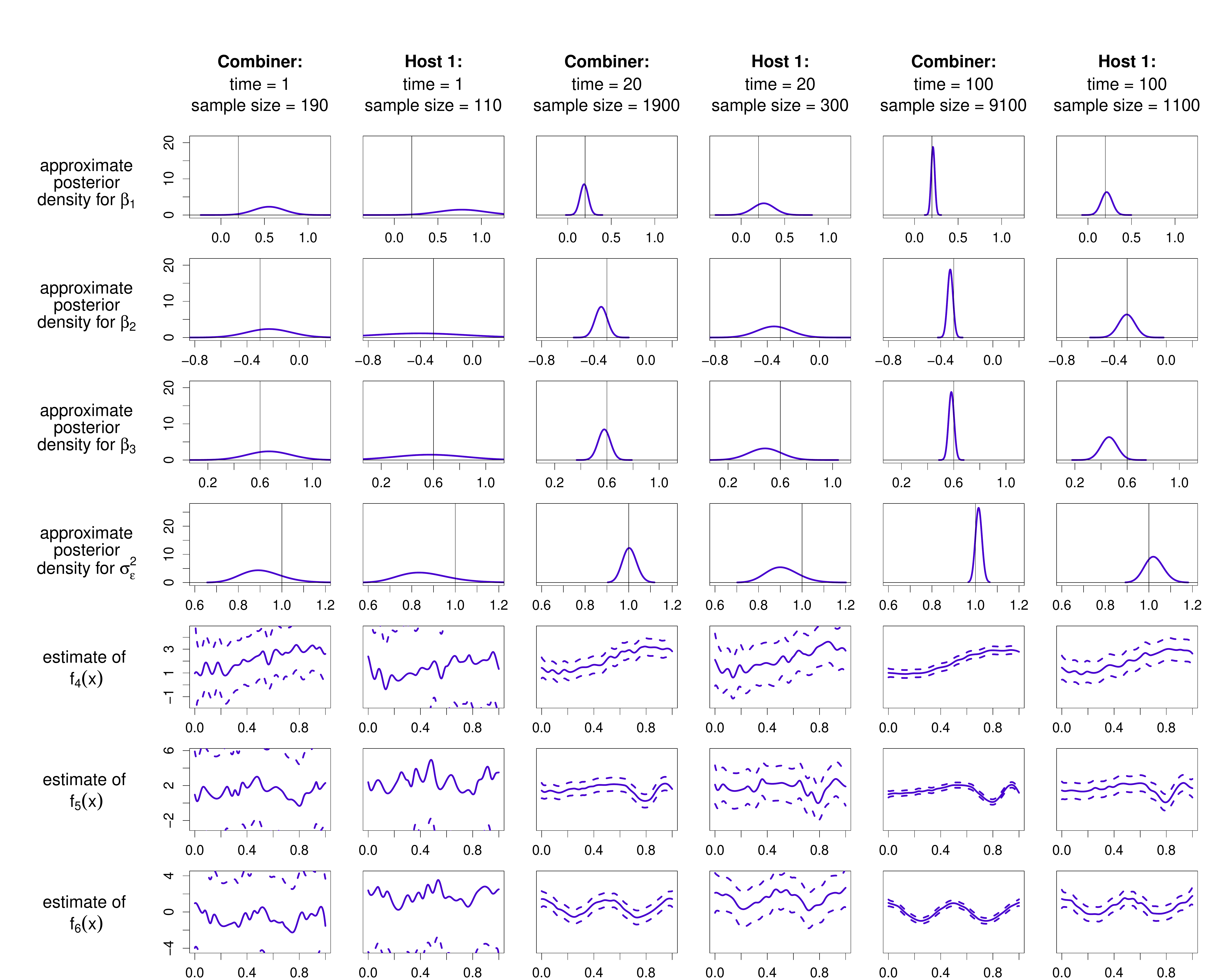}}
\caption{\it Successive approximate posterior density functions for regression coefficients and regression fits (solid lines) at the combiner and at host 1 out of 9 for the synthetic data example in Section \ref{section3:b2}. Dashed lines represent corresponding 95\% credible sets and the number of samples that has been processed at the combiner and host 1 at time = 1, 20 and 100 is indicated at the top. These results are based on Algorithm \ref{algDistOnline} for real-time distributed MFVB. The axis limits are the same across each row and a vertical line is positioned at the true value.}
\label{fig:realTimeDistributed} 
\end{figure}

Consider the following model for a synthetic data example to illustrate Algorithm \ref{algDistOnline},

\begin{equation}
y_i|\bbeta,\bu_4,\bu_5,\bu_6,\sigeps^2 \simind N\Big(\beta_1\,x_{1i}+\beta_2\,x_{2i}+\beta_3\,x_{3i} +f_4(x_{4i})+f_5(x_{5i})+f_6(x_{6i}),\sigeps^2\Big), \nonumber \\
\end{equation}

\noindent where $\bu_\ell$ is the vector of spline coefficients for $f_\ell(\cdot)$ with $\ell=4,5$ and $6$. The number of hosts is fixed at $h=9$ and each of these hosts processes 1000 samples, generated according to the model above with $x_{1i},x_{2i},x_{3i}\simind\mbox{Bernoulli}\,(1/2)$ and $x_{4i},x_{5i},x_{6i}\simind N(0,1)$. The true values were set at $\beta_1=0.2$, $\beta_2=-0.3$, $\beta_3=0.6$, $f_4(x)=2\Phi(6x-3)$, $f_5(x)=\sin(3\pi x^3)$, $f_6(x)=\cos(4\pi x)$ and $\sigeps^2=1$. Each host individually processes its incoming data in an online manner and, when its local buffer contains summary statistics from 10 samples, it sends the corresponding sums to the combiner. Assuming that all hosts simultaneously process their data at the same rate, the combiner receives summary statistics from 90 samples at each time instance. Figure \ref{fig:realTimeDistributed} visualizes the approximate posterior density functions for the regression coefficients and the regression fits at the site of the combiner and host 1. The approximate posterior density functions are first, i.e. time = 1, flat at host 1 and the combiner and regression fits show noisy relationships. As time progresses, i.e. time = 20 and 100, these regression summaries start to approximate the true underlying values and relationships. The link \texttt{Real-time Gaussian additive model for distributed data} on the website \texttt{realtime-semiparametric-regression.net} points to a movie showing summaries of the regression fits when the streaming data are simultaneously processed at 9 hosts and the combiner. Convergence to the true values and nonlinear relationships is faster at the combiner than at an individual host, illustrating the power of a real-time distributed semiparametric regression analysis. The difference in rate of convergence is dependent on the number of hosts $h$. Note that a warm-up sample of size 100 was used at the combiner and all 9 hosts, providing starting values for $\mu_{q(1/\sigeps^2)}$, $\mu_{q(1/\sigma_{u\ell}^2)}$, $\bC^T\bC$, $\bC^T\by$, $\by^T\by$ and $n$. This explains the sample sizes $110$ and $190$ for host 1 and the combiner at time = 1, respectively. 

\section{Evolving environments} \label{section4}

The previous section outlined an algorithm for continuous learning for distributed data sets based on the assumption that the underlying true model (e.g. $\bbeta$, $\bu$ and $\sigeps^2$) does not change over time. In this section, two approaches are proposed to relax this assumption since handling evolving environments is almost inherently connected with real-time streaming data analysis: the characteristics of the new incoming data can change over time in a data stream. The first approach relies on the definition of a time window, while the second is based on reweighting older data.

\subsection{Adaptation through a time window} \label{section4a}

When data arrive in a stream, newer samples are often more relevant for the present situation than older samples. For example, housing market data from the last month might be more informative than data from 24 months ago if one aims to create a predictive model for the near future. However, in some situations data from the same month, season (or quarter in economics) from the previous year might be more relevant than the previous month or season of the current year. In both situations it is often possible to define an appropriate time period of interest, such that only samples from within that specific time frame contribute to the regression fit. In this paper the time period of interest is called the time window and, as time evolves, the time window is shifted so that older samples leave the window and new samples enter the window. Although this section assumes a fixed window width, the methodology can be generalized to a time-variable window width.

Extending Algorithm \ref{algDistOnline} towards evolving environments using a time window simply requires modifying lines 4--5 to

\begin{equation}
\bC^T\bC \leftarrow \bC^T\bC + \sum_{b=1}^{B}\bC_b^T\bC_b - \bC_{\text{old}}^T\bC_{\text{old}}; \quad \bC^T\by \leftarrow \bC^T\by + \sum_{b=1}^{B} \bC_b^T\by_b - \bC_{\text{old}}^T\by_{\text{old}} 
\label{window1}
\end{equation}

\noindent and

\begin{equation}
\by^T\by \leftarrow \by^T\by + \sum_{b=1}^{B} \by_b^T\by_b -  \by_{\text{old}}^T\by_{\text{old}}; \quad n \leftarrow n + \sum_{b=1}^{B} n_{\text{old}} - n_{\text{old}} 
\label{window2}
\end{equation}

\noindent where $\bC_{\text{old}}$, $\by_{\text{old}}$ and $n_{\text{old}}$ correspond to the data and number of samples that leave the time window at a certain point in time, respectively. Note that this extension, in contrast to the original Algorithm \ref{algDistOnline} in Section \ref{section3:b}, assumes that new data are temporally stored such that their contribution to the summary statistics can later be removed. 

For illustrative purposes, the first synthetic data example in this section considers the simple linear regression model,

\begin{equation}
y_i|\beta_0,\beta_1,\sigeps^2 \simind N\Big(\beta_0+\beta_1\,x_{i},\sigeps^2\Big)\quad \text{with} \quad x_i \simind \text{Uniform}\left(0,1\right). \\
\label{generatedFromLinearRegression}
\end{equation}

\begin{figure}[!ht]
\centering
{\includegraphics[width=\textwidth]{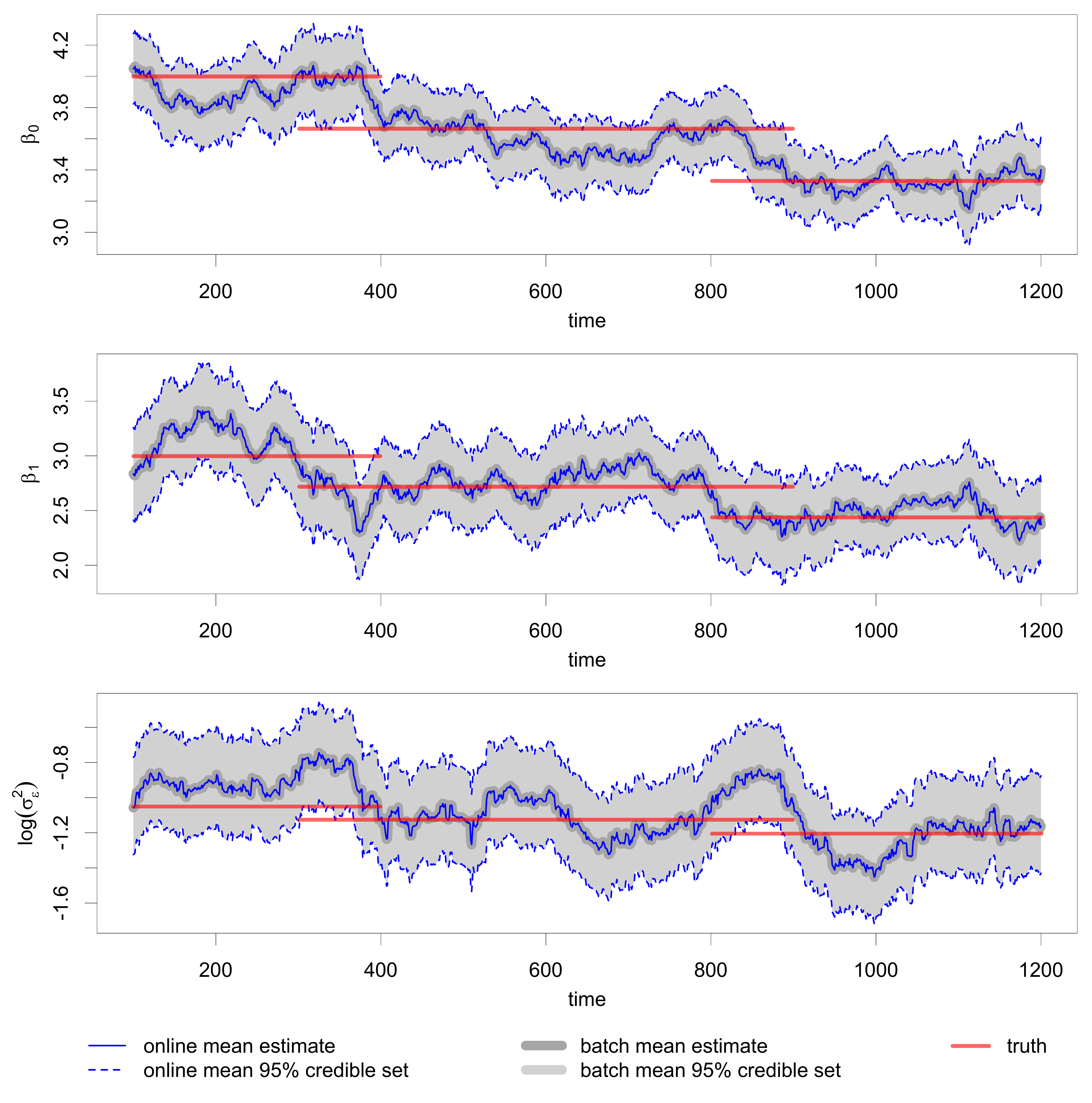}}
\caption{\it Evolution over time of the online and batch mean estimates and 95\% credible sets for a time-evolving data set generated according to the simple linear regression model in (\ref{generatedFromLinearRegression}). The online approach based on a time window of 100 samples produces similar results as compared to repeatedly executing a batch algorithm and it captures the evolving, true regression relationship (horizontal red lines) over time.}
\label{fig:forgetWindowLin} 
\end{figure}

\noindent The true values for the different parameters are gradually decreased over time: $\beta_0 \in \left\{4,3.665,3.33\right\}$, $\beta_1 \in \left\{3,2.72,2.44\right\}$ and $\sigeps^2 \in \left\{0.350,0.325,0.300\right\}$. For the first, second and third combination of parameters 300, 500 and 400 $(x_i,y_i)$-pairs were generated, respectively. Figure \ref{fig:forgetWindowLin} visualizes the result from applying a simplified version (i.e. the linear regression model above being a simplification of model (\ref{lmm})) of Algorithm \ref{algDistOnline} with the extension in (\ref{window1})--(\ref{window2}) to this data set. A fixed window size of 100 samples is used and the $i$th time instance on the horizontal axis represents the moment when the $i$th and $(i-100)$th sample enters and leaves the time window, respectively. For example, the estimated value for $\beta_0$ at time = 200 is exclusively based on samples 101 to 200. The online algorithm is compared with batch Algorithm \ref{algDistBatch} which is used on all samples in the current time window. Each time an old (new) sample leaves (enters) the time window the full batch analysis needs to be repeated entirely. The results show that the mean estimates and 95\% credible sets from the online algorithm and the batch algorithm coincide. In addition, the underlying truth is contained in the 95\% credible sets for all parameters. In case the time window starts to contain samples being generated from different $\bbeta$ or $\sigeps^2$ values (i.e. when the red horizontal lines overlap in time), the estimates tend to enter a transition phase between the true, underlying values after which stable estimates are again obtained. 

\begin{figure}[!ht]
\centering
{\includegraphics[width=\textwidth]{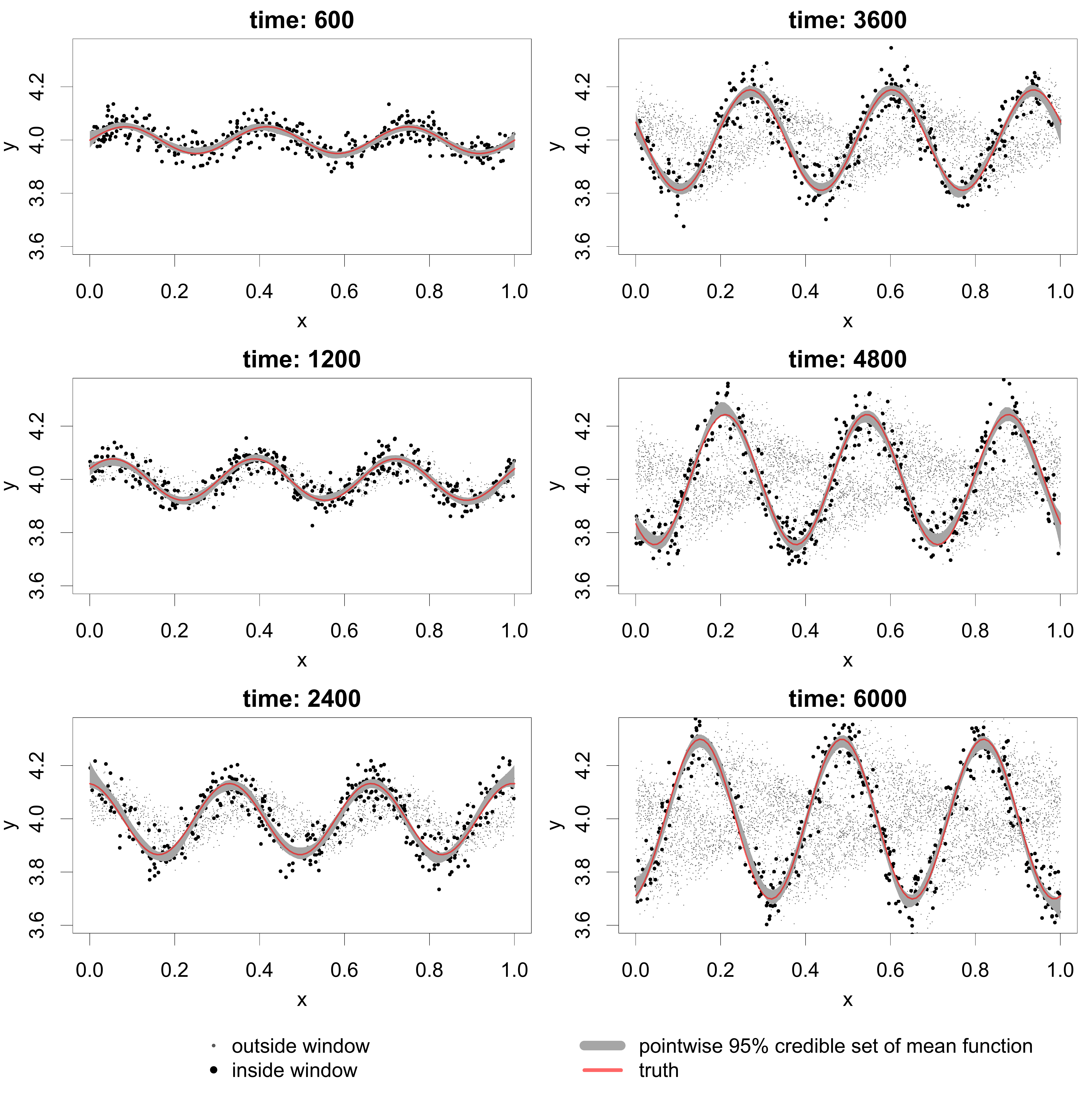}}
\caption{\it Evolution over time of the online estimate, using a time window of size 400, of the 95\% pointwise credible set of the mean function for a time-evolving data set generated according to the model in (\ref{generatedFromMixedRegression}). Black dots represent samples that are currently in the time window, while samples outside the window are visualized as small grey dots. The true nonlinear relationship is superimposed as a red curve, illustrating that the regression method captures the evolving nonlinear relationship.}
\label{fig:forgetWindowMixed} 
\end{figure}

The next example fits a semiparametric regression model on synthetic data that were generated according to 

\begin{equation}
y_i|\alpha_0,\alpha_1,\alpha_2,\sigeps^2 \simind N\Big(\alpha_0+\alpha_1\,\sin(6\pi x_{i}+\alpha_2),\sigeps^2\Big) \quad \text{and} \quad x_i \simind \text{Uniform}\left(0,1\right), \\
\label{generatedFromMixedRegression}
\end{equation}

\noindent with $\alpha_0=4$, $\alpha_1 \in \left\{0.5,\ldots,3\right\}$, $\alpha_2 \in \left\{0,\ldots,5 \right\}$ and $\sigeps^2 \in \left\{0.1,\ldots,0.4 \right\}$. The values for $\alpha_1$, $\alpha_2$ and $\sigeps^2$ gradually evolve in 10 equally spaced steps between these boundaries and for each combination 600 $(x_i,y_i)$-pairs were generated. Figure \ref{fig:forgetWindowMixed} visualizes 95\% credible sets for the mean at 6 different time points for this data set. Samples within the time window of size 400 are denoted by black dots while older (i.e. outside the time window) data are indicated by small grey dots. Comparing the 95\% credible sets with the true, underlying model (i.e. red curve) at each time point shows that the estimates capture the evolving true nonlinear relationship.  

\subsection{Forgetting by reweighting data}

The methodology from Section \ref{section4a} represents an appropriate approach when one is interested in real-time modeling using data from a predefined period of time. For example, one aims to continuously update a regression model such that only (distributed) data from the last 30 days contribute to the fit. However, a potential issue with the method in the previous section is that the summary statistics from data need to be stored until those data fall out of the time window. This section presents a different approach for incorporating a mechanism that enables real-time semiparametric regression for evolving environments without having to store the data or corresponding individual summary statistics.

Closer inspection of Algorithm \ref{algDistOnline} for online MFVB semiparametric regression of distributed data reveals that all summary statistics contribute with equal weight to the total sums (cf. line 4 and 5). In order to forget older information and focus on more recent data Algorithm \ref{algDistOnlineReweighting} uses a decaying window through the introduction of reweighting for the summary statistics. For example, reweighting for the summary statistic $\by^T\by$ at time $t$ can be imposed via the assignment

\begin{equation}
\by^T\by \leftarrow (1-\rho_t)\by^T\by +\rho_t \sum_{b=1}^{B} \by_b^T\by_b,
\end{equation}

\noindent where $\rho_t$ denotes the learning rate at time $t$. Various ways exist to define the learning rate: it can be kept constant over time or an adaptive approach can be used. An example of a decreasing learning rate is $\rho_t=(\tau+t)^{-\kappa}$, with fixed parameters $\tau>0$ and $\kappa \in \{k \in \mathbb{R}\,|\,0.5<k\leq 1\}$. With this decreasing learning rate, larger values for $\tau$ and $\kappa$ result in less forgetting of older samples. In addition, the level of forgetting is decreased as time evolves (i.e. for increasing $t$). On the other hand, a constant learning rate can be used to impose a constant level of forgetting over time. Similarly to the time window approach from the previous section, a constant learning rate enables us to specify that only a fixed number of most recent samples contribute to the sum of summary statistics. Whereas the samples in the time window have an equal contribution to the sum, the constant learning rate approach implies an additional weighting such that the most recent samples have higher weights. In this way the assumption that samples need to be stored can be omitted. Note that Algorithm \ref{algDistOnlineReweighting} incorporates a decreasing learning rate $\rho_t$. Using a constant learning rate simply requires to fix $\rho_t = \rho$ beforehand, where $0<\rho<1$.

\vspace{0.3cm}
\begin{algorithm}
\noindent \caption{\textit{Online mean field variational Bayes algorithm for distributed data sets with reweighting of old samples.}}
\label{algDistOnlineReweighting}
\begin{algorithmic}[1]
\REQUIRE $p, P, K_{\ell}, \mu_{q(1/\sigeps^2)}, \Aeps, \mu_{q(1/\sigma_{u\ell}^2)}, A_{u\ell}, \sigma_{\beta}^{2}, \tau, \kappa \, \, \text{with} \, 1\le\ell\le r$
\STATE $\bC^T\bC \leftarrow \bzero_{P \times P}; \quad \bC^T\by \leftarrow \bzero_{P \times 1}; \quad \by^T\by \leftarrow 0; \quad n \leftarrow 0; \quad t \leftarrow 0$
\WHILE{new data available in buffer}
		\STATE $\text{retrieve and remove} \, \, \sum_{b=1}^{B}\bC_b^T\bC_b, \, \, \sum_{b=1}^{B} \bC_b^T\by_b, \, \, \sum_{b=1}^{B} \by_b^T\by_b \, \, \text{and} \, \, \sum_{b=1}^{B} n_b \, \, \text{from buffer}$
		\STATE $t \leftarrow t + 1; \quad \rho_t \leftarrow (\tau+t)^{-\kappa}$
		\STATE $\bC^T\bC \leftarrow (1-\rho_t) \bC^T\bC + \rho_t \sum_{b=1}^{B}\bC_b^T\bC_b; \quad \bC^T\by \leftarrow (1-\rho_t) \bC^T\by + \rho_t \sum_{b=1}^{B} \bC_b^T\by_b$
		\STATE $\by^T\by \leftarrow (1-\rho_t)\by^T\by +\rho_t \sum_{b=1}^{B} \by_b^T\by_b; \quad n \leftarrow n + \sum_{b=1}^{B} n_b; \quad \gamma \leftarrow n/\{\sum_{b=1}^{B} n_b\}$
		\STATE $\bSigma_{q(\bbeta,\bu)}\leftarrow \left[\mu_{q(1/\sigeps^2)}\,\gamma\,\bC^T\bC+ \mbox{blockdiag}\{\sigma_{\beta}^{-2}\,\bI_p, \mu_{q(1/\sigma_{u1}^2)}\bI_{K_1},\ldots, \mu_{q(1/\sigma_{ur}^2)}\bI_{K_r}\} \right]^{-1}$    
   	\STATE $\bmu_{q(\bbeta,\bu)}\leftarrow 
\mu_{q(1/\sigeps^2)}\,\bSigma_{q(\bbeta,\bu)}\,\gamma\,\bC^T\by;\quad \mu_{q(1/\aeps)}\leftarrow 1/\{\mu_{q(1/\sigeps^2)}+\Aeps^{-2}\}$
		\STATE $\mu_{q(1/\sigeps^2)}\leftarrow 
\displaystyle{\frac{n+1}{2\,\mu_{q(1/\aeps)}
+\gamma(\by^T\by-2\bmu_{q(\bbeta,\bu)}^T\bC^T\by
+\mbox{tr}[(\bC^T\bC)\{\bSigma_{q(\bbeta,\bu)}+\bmu_{q(\bbeta,\bu)}
\bmu_{q(\bbeta,\bu)}^T\}])}}$
		\FOR{$\ell = 1 \to r$}
				\STATE $\mu_{q(1/a_{u\ell})}\leftarrow 1/\{\mu_{q(1/\sigma_{u\ell}^2)}+A_{u\ell}^{-2}\};\quad \mu_{q(1/\sigma_{u\ell}^2)}\leftarrow 
\displaystyle{\frac{K_{\ell}+1}{2\,\mu_{q(1/\auell)}
+\Vert\bmu_{q(\bu_{\ell})}\Vert^2
+\mbox{tr}(\bSigma_{q(\bu_{\ell})})}}$
		\ENDFOR
\ENDWHILE
\end{algorithmic}
\end{algorithm}
\vspace{0.3cm}

\begin{figure}[!ht]
\centering
{\includegraphics[trim = 13mm 20mm 230mm 15mm, clip, width=\textwidth]{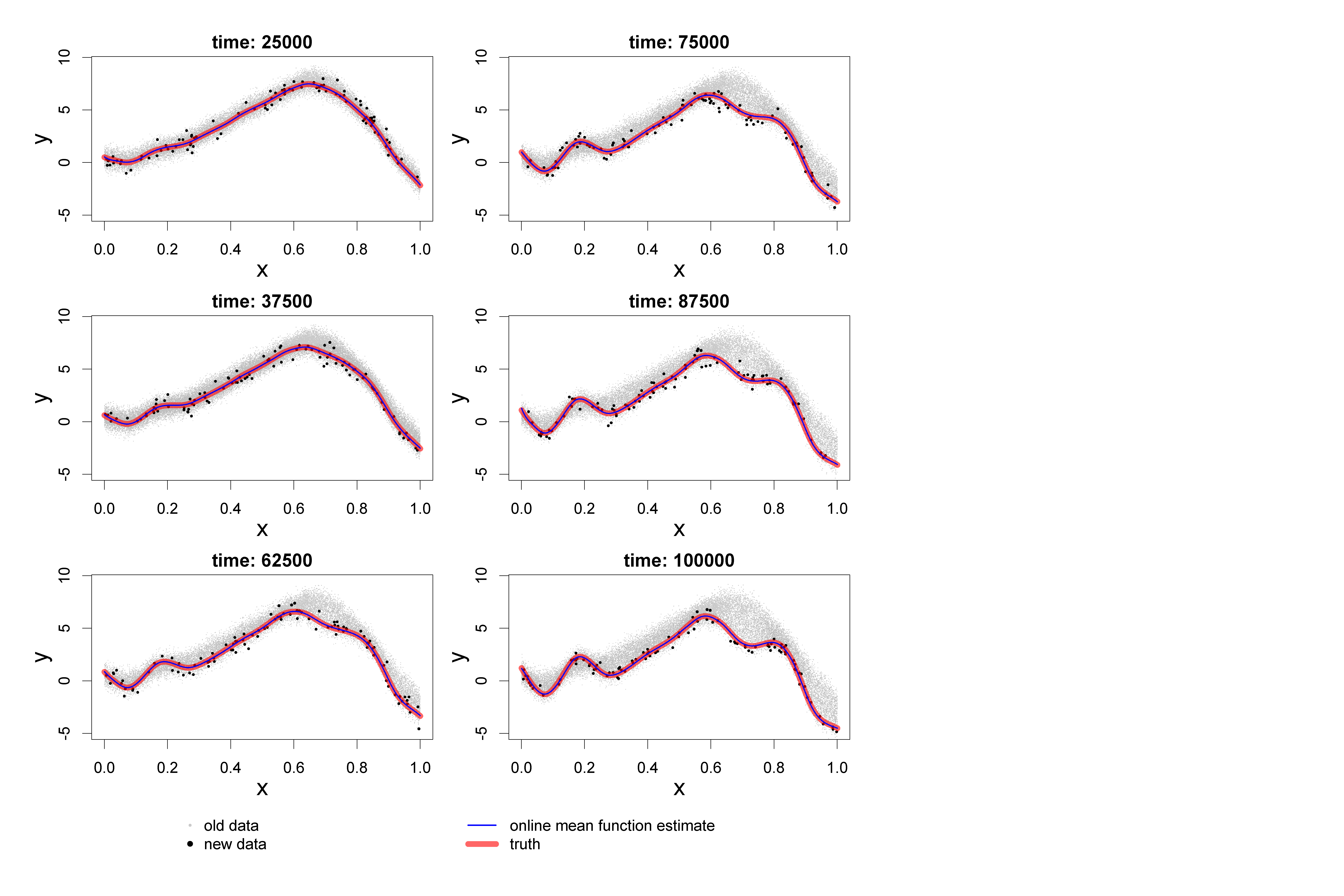}}
\caption{\it Evolution over time of the online estimate of the mean function, obtained using Algorithm \ref{algDistOnlineReweighting}, for a time-evolving data set generated according to the model in (\ref{generatedFromMixedRegressionWeighting}). The true mean is displayed as a red curve, while the estimated mean is denoted by the thin blue line. The 100 most recent data points are visualized as black dots, while small grey dots represent older samples. Reweighting summary statistics using a constant learning rate $\rho_t=0.001$ enables to model the true nonlinear relationship over time.}
\label{fig:forgetReweightingMixed} 
\end{figure}

To illustrate Algorithm \ref{algDistOnlineReweighting}, data were generated according to 

\begin{equation}
\begin{array}{c}
y_i|\beta_0,\beta_1,u_1, \ldots , u_{24} \simind N\left(\beta_0+\beta_1 x_i+\sum_{k=1}^{24} u_k z_k(x_i),0.25\right),\quad x_i \simind \text{Uniform}\left(0,1\right), \\ 
\vspace{-0.4cm}
\end{array}
\label{generatedFromMixedRegressionWeighting}
\vspace{0.4cm}
\end{equation}

\noindent where the true values for $\beta_0,\beta_1,u_1, \ldots , u_{24}$ were gradually modified using linear interpolation. The number of data hosts, i.e. $B$, was fixed at 10 and $n_b=1$ was kept constant. At each time instance 10 samples were processed and the total number of time instances equaled 100000. The true values for the model parameters were modified each 12500 time instances. Figure \ref{fig:forgetReweightingMixed} displays the evolution of the true relationship between the independent and response variable as a red curve at six time points. The 100 most recent samples are plotted as black dots while older data are visualized as small grey dots. The thin blue curve shows the estimate of the mean by using Algorithm \ref{algDistOnlineReweighting} with a fixed learning rate $\rho_t=0.001$. Figure \ref{fig:forgetReweightingMixed} shows that the estimated mean adapts itself to the time-evolving data.

\section{Application areas} \label{section5}

The algorithms presented in the previous sections have assumed that there exists one combiner that merges all the contributions from the individual data hosts to obtain the global regression result. This setting is potentially useful for a wide range of applications. For example, a large retail company with several local stores wishes to analyze consumer behavior and local store performance in real-time. Each local store collects data about individual consumer purchases every few seconds and corresponding local summary statistics are continuously combined in the analytics department of the company. Algorithm \ref{algDistOnlineReweighting} allows us to handle such grouped (e.g. the local store and/or the individual consumer) data: patterns of behavior over time can be interpreted and the performance of each individual store can be monitored in real-time for presentation to management.

This section further explains other possible scenarios for the use of the proposed algorithms. The first example deals with the situation where there exist multiple data owners that want to do cooperative semi-parametric regression, but without disclosure of their data or summary statistics. The second example addresses the use of the algorithms within the context of the MapReduce programming model for distributed computing. 

\subsection{Multiple data owners: cooperative analysis} \label{section5:a}

When mutually untrusted parties or competitors jointly aim to conduct the proposed semi-parametric regression analysis, privacy becomes an important issue. Even though the various parties are likely to benefit from a cooperative analysis, their highest priority might still be protecting the confidentiality of their own data. For example, the parties might not be willing to share individual records, nor to reveal the origin of the data. Assuming the existence of a trusted third party that performs the analysis is not always realistic, secure multiparty computation has a role to play \citep{DU:2001}.

The algorithms that were presented in the previous sections do not require sharing individual data records (i.e. samples), but are based on sharing summary statistics. Secure multiparty computation in such a context requires a method for secure summation of these summary statistics. The outcome of such a secure summation is that the different parties, or data hosts as described in Algorithm \ref{algDistBatch}, obtain the combined results, i.e. $\bC^T\bC, \bC^T\by, \by^T\by$ and $n$, but gain no information about the individual summary statistics of the other parties. This includes both protecting the summary statistics and their origin. A simple secure summation protocol to compute $n$ for $B>2$ parties is as follows:

\begin{itemize} 

\item Party 1 generates a large random integer $n_{\text{random}}$ and sends $n_{\text{random}}+n_1$ to Party 2.
\item Party 2 adds $n_2$ to the input it received from Party 1 and sends the result to Party 3, etc.
\item Party 1 subtracts $n_{\text{random}}$ from the number it received from Party $B$ and shares the result with all other parties.  

\end{itemize}

An identical protocol can be followed to compute $\by^T\by$ and, similarly, it can be used to compute the (unique) entries of $\bC^T\bC$ and $\bC^T\by$. Remark that this secure summation protocol assumes that the different parties correctly follow the protocol specification and that they use their true data. \citet{GHOSH:2007} used such a protocol for secure multiparty computation for multivariate adaptive regression splines to model nonlinear relationships. Compared to \citet{GHOSH:2007} the Bayesian penalized splines approach in this paper offers the advantage of automated regularization parameter inference, providing measures of uncertainty (e.g. credible sets) and extensions to more complicated graphical models (e.g. grouped data, geostatistical data or sparse priors) are straightforward. In addition, online (cf. Algorithm \ref{algDistOnline}) instead of batch computation can also be used in a secure multiparty computation context. However, this requires repeatedly applying the protocol above, which might be time-consuming. 

Alternatively, a network-based client-server model can be used for secure computation as in \citet{KARR:2007}, where it was used for linear regression. In this way, the parties do not directly interact with each other but only through a server, having the advantage of randomizing the order in which messages are sent between the clients, i.e. parties. Encryption technology prevents the server from actually reading the summary statistics it passes between the clients.

\subsection{MapReduce for processing large data sets} \label{section5:b}

The MapReduce programming model was developed at the Internet technology company Google Inc. for distributed processing of very large data sets \citep{DEAN:2008}. Being confronted with huge computing tasks Google Inc. decided to take advantage of a distributed computing environment, where large clusters of hundreds or thousands of commodity computers are connected together. Such a setting requires a system for taking care of partitioning the input data, scheduling the execution across the commodity computers, handling failure of computers and managing communication between the machines. The MapReduce framework provides a convenient way to handle these tasks and enables programmers without any experience with parallel and distributed systems to make use of the resources of distributed processing. In essence, MapReduce can be used in conjunction with various architectures. For example, \citet{CHU:2007} presented a MapReduce implementation based on multicore computers, thereby taking advantage of the shared memory. In this section no assumptions are made about the underlying architecture as the main aim is to provide the general flavor of how the proposed algorithms fit into the MapReduce programming paradigm. In addition, the issue of when to opt for MapReduce over another distributed system is out of the scope of this paper.

A MapReduce task consists of a map phase and a reduce phase and the user has to specify the corresponding map and reduce function. The map function processes key-value pairs and outputs intermediate key-value pairs. Typically, the map task can be distributed over multiple machines, each operating in parallel on a small subset of the total data set. The reduce function then processes all the intermediate values that share the same intermediate key and outputs the final result. Essentially, the reduce function combines the intermediate results from the map function. Optionally, there is the possibility to implement a combiner function, which operates before the reduce phase starts. The
combiner function is typically identical to the reduce function, but it is executed on each computer that performs a map task. This has the advantage of speeding up the computations when there exists significant repetition in the intermediate keys.

A factor that strongly popularized the use of MapReduce, was the development of an open-source implementation called Hadoop \citep{WHITE:2009}. While Hadoop was directly derived from Google Inc.'s MapReduce and the Google File System, a number of related projects have emerged in recent years. For example, the Mahout project is concerned with free implementations of distributed or otherwise scalable machine learning algorithms on the Hadoop platform (\texttt{http://mahout.apache.org/}). Most of the algorithms that are implemented in Mahout are concerned with clustering or classification. Here, a map and reduce function are presented for batch semiparametric regression when the data are stored in a distributed file system (Algorithm \ref{algMapReduce}). The map function basically computes the summary statistics (as needed for executing Algorithm \ref{algDistBatch}) based on subset $D_i$ of the total data set. Firstly, $\bC, \by$ and $n$ are extracted from $D_i$ and then the summary statistics based on this subset of samples are emitted together with the corresponding intermediate key. The reduce function simply sums the summary statistics with the same intermediate key together and outputs this result. Finally, lines 6--13 of Algorithm \ref{algDistBatch} are used to compute $\bSigma_{q(\bbeta,\bu)}$ and $\bmu_{q(\bbeta,\bu)}$. Note that although lines 6--13 are iterative, the individual steps can again be parallelized (e.g. computing $\bSigma_{q(\bbeta,\bu)}$ or computing $\mu_{q(1/a_{u\ell})}$ and $\mu_{q(1/\sigma_{u\ell}^2)}$ for different values of $\ell$) as explained in Section \ref{section3:b}.

Although MapReduce was originally developed for computing batch jobs, a lot of research is going on to adapt it to process data streams. \citet{CONDIE:2010} proposed a modified MapReduce architecture called Online MapReduce, that allows mappers to push data to reducers as soon as it is generated. This type of pipelining between mappers and reducers enables online aggregation and continuous queries. Online aggregation means that an intermediate result is generated during the course of execution, instead of having to wait for obtaining the final result till the job is completely finished. In addition, their architecture allows for real-time processing of data streams: the Online MapReduce jobs can run continuously, accept new data as it becomes available and process it immediately. Algorithm \ref{algDistOnline} for real-time semiparametric regression of distributed data sets nicely fits into this architecture. By using the map and reduce function from Algorithm \ref{algMapReduce}, Online MapReduce repeatedly generates updated summary statistics. Each time the updated summary statistics are being outputted, lines 6--10 of Algorithm \ref{algDistOnline} are executed, continuously producing estimates for $\bSigma_{q(\bbeta,\bu)}$ and $\bmu_{q(\bbeta,\bu)}$.

\vspace{0.3cm}
\begin{algorithm}
\noindent \caption{\textit{Map and reduce function for computing the summary statistics as used in batch mean field variational Bayes Algorithm \ref{algDistBatch}.}}
\label{algMapReduce}
\begin{algorithmic}[1]
\STATE \textbf{function} map(key $i$, data set $D_i$)
		\STATE \hspace{0.3cm} extract $\bC, \by$ and $n$ from data set $D_i$
		\STATE \hspace{0.3cm} compute $\bC^T\bC, \bC^T\by$ and $\by^T\by$
		\STATE \hspace{0.3cm} \textbf{emit}(1,$\bC^T\bC$)		
		\STATE \hspace{0.3cm} \textbf{emit}(2,$\bC^T\by$)		
		\STATE \hspace{0.3cm} \textbf{emit}(3,$\by^T\by$)					
		\STATE \hspace{0.3cm} \textbf{emit}(4,$n$)
\vspace{0.5cm}		
\makeatletter\setcounter{ALC@line}{0}\makeatother
\STATE \textbf{function} reduce(key $i$, list $L$)
		\STATE \hspace{0.3cm} $S \leftarrow$ sum($L$)
		\STATE \hspace{0.3cm} \textbf{emit}($i$,$S$)
\end{algorithmic}
\end{algorithm}
\vspace{0.3cm}

\section{Example: real-time processing of U.S. domestic flight data} \label{section6}

This section demonstrates the proposed methodology by processing U.S. domestic flight data with the goal to analyze air traffic delays in real time. Nowadays, the status of a flight is continuously monitored and airports generate data streams which contain information about, among other things, the actual runway and gate arrival times for thousands of flights per day. For this example, the website \texttt{www.flightstats.com} is used to get access to these real-time data on flight delays, flight distances, operating airlines and flight paths. In addition, air temperature, wind speed and aviation flight category observations are continuously made at airports and nearby weather observation stations. These weather reports can be produced by automated airport weather stations or by trained observers or forecasters who manually observe and encode their observations. Here, these data are obtained through the \texttt{aviationweather.gov} website.

In this example the real-time flight data consist of the flight delay, flight distance, operating airline and the flight path. The real-time weather data consist of air temperature, wind speed and aviation flight category measurements. The flight category is a combined measure for the visibility and ceiling and there exist four categories: visual flight rules (VFR, i.e. ceiling $>$ 3000 feet and visibility $>$ 5 miles), marginal visual flight rules (MVFR, i.e. 1000 feet $\leq$ ceiling $\leq$ 3000 feet and/or 3 miles $\leq$ visibility $\leq$ 5 miles), instrument flight rules (IFR, i.e. 500 feet $\leq$ ceiling $<$ 1000 feet and/or 1 mile $\leq$ visibility $<$ 3 miles) and low instrument flight rules (LIFR, i.e. ceiling $<$ 500 feet and/or visibility $<$ 1 mile). 

An extension of semiparametric regression model (\ref{eq:randInt}) is used to demonstrate the methodology:
\begin{equation}
\begin{array}{l}
\log(\mbox{\texttt{delay}}_{ijk}+120)|\,\bbeta,U_i, V_j,\bu_7,
\bu_8,\bu_9,\bu_{10},\bu_{11},\sigeps^2\simind \\[1ex]
\hspace{2mm} N(\beta_0 + \beta_1 {\mbox{\texttt{MVFRdep}}}_{ijk} 
+ \beta_2 {\mbox{\texttt{IFRdep}}}_{ijk} + \beta_3 {\mbox{\texttt{LIFRdep}}}_{ijk}  
+ \beta_4 {\mbox{\texttt{MVFRarr}}}_{ijk}\\[1ex]
\hspace{2mm} + \beta_5 {\mbox{\texttt{IFRarr}}}_{ijk} 
+ \beta_6 {\mbox{\texttt{LIFRarr}}}_{ijk}  + f_7({\mbox{\texttt{flight distance}}}_{j} )
\\[1ex]
\hspace{2mm} + f_8({\mbox{\texttt{departure temperature}}}_{ijk} ) 
+ f_9({\mbox{\texttt{arrival temperature}}}_{ijk} )
\null\\[1ex]
\hspace{2mm} +  f_{10}({\mbox{\texttt{departure wind speed}}}_{ijk} ) 
+ f_{11}({\mbox{\texttt{arrival wind speed}}}_{ijk} )
\null\\[1ex]
\hspace{2mm} + U_i + V_j,\sigeps^2),
\hspace{3mm} U_1,\hdots,U_{171} |\,\sigma^2_{U} \simind N(0,\sigma^2_{U}),
\hspace{3mm} V_1,\hdots,V_{2000} |\,\sigma^2_{V} \simind N(0,\sigma^2_{V}).
\end{array}
\label{eq:DomesticFlightData}
\end{equation}

\noindent Here, \texttt{delay}$_{ijk}$ is the difference between the actual and scheduled runway arrival time in minutes for the $k$th flight of airline $i$ on flight path $j$. \texttt{MVFRdep}$_{ijk}$ is an indicator which equals 1 if MVFR are applied at the scheduled runway departure time of the $k$th flight of airline $i$ on flight path $j$ and 0 otherwise. \noindent The variable $\texttt{MVFRarr}_{ijk}$ is defined in an analogous way, but for the scheduled runway arrival time. The other aviation flight category variables are defined similarly. The variable \texttt{flight} \texttt{distance}$_{j}$ represents the distance of flight path $j$ in kilometers. Variables \texttt{departure temperature}$_{ijk}$ and \texttt{arrival temperature}$_{ijk}$ denote the air temperature in degrees Celsius at the scheduled runway departure and arrival time of the $k$th flight of airline $i$ on flight path $j$, respectively. Variables \texttt{departure wind speed}$_{ijk}$ and \texttt{arrival wind speed}$_{ijk}$ denote the wind speed in 
knots at the scheduled runway departure and arrival time of the $k$th flight of airline $i$ on flight path $j$, respectively. Random intercepts for each of the 171 airlines are denoted by $U_i$,$\, 1 \leq i \leq 171$, and random intercepts for each of the 2000 flight paths are defined by $V_j$,$\, 1\leq j \leq$ 2000. $\bbeta$ stores the fixed effect regression coefficients $\beta_0,\hdots,\beta_6$ and the linear contribution to $f_7, \hdots , f_{11}$. The spline basis coefficients for $f_7, \hdots , f_{11}$ are stored in $\bu_7, \hdots ,\bu_{11}$.

Algorithm \ref{algDistOnline} is used to fit model (\ref{eq:DomesticFlightData}) and the time window extension in assignments (\ref{window1})--(\ref{window2}) is implemented to focus only on the 30 most recent days, i.e. a time window of 30 days is used. The website \texttt{realtime-semiparametric-regression.net/FlightDataForgetting/} demonstrates fitting of 
(\ref{eq:DomesticFlightData}) using this methodology and presents continuously updated results in real time. To highlight the advantage of distributed processing through Algorithm \ref{algDistOnline} the combined results for data generated by 415 U.S. airports (i.e. the data hosts) are presented together with the separate results, obtained by independently using Algorithm 3 of \citet{LUTS:2013} extended with time window assignments (\ref{window1})--(\ref{window2}), for \texttt{O'Hare International Airport} and \texttt{Dallas/Fort Worth International Airport}. The first table shows the influence of flight distance, airline and the weather at the departure and arrival airport on the flight delay using summary statistics from all airports, from only \texttt{O'Hare International Airport} and from only \texttt{Dallas/Fort Worth International Airport}. Particularly interesting are the top 10 airlines having lowest and highest delays during the last 30 days after correcting for all other covariates such as weather circumstances and airports through the flight path variable. Similarly, the second table provides the top 10 flight paths having lowest and highest delays during the last 30 days. All these regression summaries are computed in real-time and the figures are updated every few minutes. The figure entitled \texttt{airline delay evolution over time} visualizes the on-time performance for the major airlines \texttt{Delta Air Lines}, \texttt{United Airlines} and \texttt{Southwest Airlines} based on the estimates for the random intercepts $U_i$ by combining summary statistics from 415 airports. Each day a new data point is added to this figure for each of these airlines.

\section{Conclusion} \label{section7}

This paper proposes methodology for semiparametric regression analysis when the samples are spread over multiple data hosts. Often it is not possible to move the raw data itself due to their large-scale nature or due to confidentiality issues. Mean field variational Bayes semiparametric regression algorithms are presented for this setting, thereby allowing data to be processed in batch or in an online manner. The key aspect of these approaches is to combine summary statistics instead of actual data. Compared to earlier work on regression for distributed data sets, this allows modeling of nonlinear relationships, enables fully-automated inference for smoothing parameters and provides measures of uncertainty. Furthermore, the presented model handles complications as grouped data and the Bayesian approach permits extensions to a wider variety of models. 

An important aspect of analyzing distributed streaming data is to adapt to changes in the target over time. Two approaches have been proposed to deal with evolving environments. The first approach uses a time window to let the real-time regression estimates only depend on the most recent samples. This requires defining the window width and storing the summary statistics belonging to the time window. The second approach uses a decaying window by reweighting the summary statistics of older data and new data to handle a changing environment. This approach requires choosing a learning rate. 

In order to illustrate the practical relevance of the proposed method, two types of application areas are discussed: semiparametric regression when there are multiple data owners requiring secure multiparty computation and the use of semiparametric regression within the MapReduce programming model. Finally, the method has been demonstrated on a real-life data set. An Internet site attached to this paper visualizes semiparametric regression analysis for infinite streams of data that are generated at 415 U.S. airports in real time. Future work includes extensions to other types of regression models as for example logistic regression or models with sparsity-inducing penalties for automated variable selection.

\section*{Acknowledgments}

This research was supported by Australian Research Council Discovery Project DP110100061. The author is grateful to Alan Huang and Matt Wand for their comments.


\bibliographystyle{model4-names}
\bibliography{distributedRTVB}

\end{document}